\RequirePackage{fix-cm}
\documentclass[smallextended]{svjour3}       
\smartqed  
\usepackage{graphicx}
\usepackage{lipsum}
\usepackage{pdfpages}
\usepackage{float}
\usepackage{caption}  
\usepackage{subcaption} 
\usepackage{nicefrac}
\usepackage{url}
\usepackage{booktabs}
\usepackage{comment}
\usepackage{color,soul} 
\usepackage{array}
\usepackage{multirow}
\usepackage{bigstrut}
\usepackage{subcaption}
\usepackage{units}
\usepackage{flushend}
\usepackage{url}
\usepackage{natbib}
\usepackage{hyperref}
\usepackage{colortbl}
\usepackage{booktabs}
\usepackage{natbib}
\usepackage{makecell}
\usepackage{mathptm}
\usepackage{tcolorbox}

\usepackage{fancyvrb} 
\usepackage{fvextra} 
\usepackage{appendix}
\usepackage{microtype}

\usepackage{tabularx}

\usepackage{pdfpages}

\usepackage{xcolor}

\usepackage{tcolorbox}
\usepackage{listings}
\usepackage{fvextra}
\DefineVerbatimEnvironment{Verbatim}{Verbatim}{breaklines=true}
\newtcolorbox{highlightbox}{
	colback=gray!20, 
	colframe=black,   
	boxsep=2pt,       
	arc=0pt,          
	boxrule=1pt       
}

\newcommand{\RQone}{RQ1: What are the perceived reasons behind the differences between ML and non-ML projects with respect to their CI practices?}
\newcommand{\RQtwo}{RQ2: How do ML practitioners manage the build duration of their CI pipeline?}
\newcommand{\RQthree}{RQ3: Are there differences when handling test coverage rates between ML and non-ML projects?}

\begin{document}

\title{Continuous Integration Practices in Machine Learning Projects: The Practitioners' Perspective
}

\titlerunning{CI Practices in ML Projects: The Practitioners' Perspective}        

\author{
    João Helis Bernardo \and 
    Daniel Alencar da Costa \and
    Filipe Roseiro Cogo \and
    Sérgio Queiróz de Medeiros \and    
    Uirá Kulesza        
}



\institute{
João Helis Bernardo, Sérgio Queiroz de Medeiros, and Uirá Kulesza  \at
Federal University of Rio Grande do Norte \\
Natal, Brazil \\              
\email{joao.helis@ifrn.edu.br, sergio.medeiros@ufrn.br, uira@dimap.ufrn.br}
\and
Daniel Alencar da Costa \at
University of Otago \\
Dunedin, New Zealand \\              
\email{danielcalencar@otago.ac.nz}      \and
Filipe Roseiro Cogo \at
Centre for Software Excellence, Huawei \\
Kingston, Canada \\                     
\email{filipe.roseiro.cogo1@huawei.com}
}

\date{Received: date / Accepted: date}

\maketitle

\begin{abstract}
Continuous Integration (CI) is a cornerstone of modern software development, promoting stability and ensuring quality. 
However, while widely adopted in traditional software projects, applying CI practices to Machine Learning (ML) projects presents distinctive characteristics. For example, our previous work revealed that ML projects often experience longer build durations and lower test coverage rates compared to their non-ML counterparts.
Building on these quantitative findings, this work investigates the underlying reasons for these distinctive characteristics through a qualitative perspective.
We surveyed 155 practitioners from 47 ML projects and conducted a thematic analysis to identify the key differences in CI adoption in the ML domain compared to non-ML domains. 
Practitioners highlighted eight key differences, including test complexity, infrastructure requirements, and build duration and stability. Common challenges mentioned by practitioners include higher project complexity, model training demands, extensive data handling, increased computational resource needs, and dependency management, all contributing to extended build durations. Furthermore, ML systems' non-deterministic nature, data dependencies, and computational constraints were identified as significant barriers to effective testing.
The key takeaway from this study is that while foundational CI principles remain valuable, ML projects require tailored approaches to address their unique challenges.
To bridge this gap, we propose a set of ML-specific CI practices, including tracking model performance metrics and prioritizing test execution within CI pipelines. Additionally, our findings highlight the importance of fostering interdisciplinary collaboration to strengthen the testing culture in ML projects. Furthermore, we emphasize the need for standardized guidelines to address key CI challenges in ML workflows, such as dependency management.
By bridging quantitative findings with practitioners' insights, this study provides a deeper understanding of the interplay between CI practices and the unique demands of ML projects, laying the groundwork for more efficient and robust CI strategies in this domain.

\end{abstract}
\section{Introduction}
\label{intro}

Machine Learning (ML) has become a cornerstone of modern software systems, driving advancements across diverse domains~\citep{washizaki2019studying, gonzalez2020state, pallathadka2023impact}. 
In this paper, we define ML projects as software initiatives that integrate ML components to enable intelligent functionalities. These projects either address domain-specific problems (i.e., ML applications or ML-enabled systems) or provide general-purpose solutions (i.e., ML tools).

As ML projects gain prominence and grow in complexity, ensuring their quality and reliability becomes increasingly challenging. 
These projects often involve computationally intensive tasks and exhibit inherent non-deterministic behavior \citep{nascimento2020software, giray2021software}, making it difficult to guarantee consistent performance and reproducibility.
Furthermore, ML projects face the unique challenge of deploying models to production, a task that can further strain development workflows~\citep{symeonidis2022mlops}.

To tackle these complexities, the field of Software Engineering for Machine Learning (\textsc{SE4ML}) has emerged, adapting traditional software engineering practices to the specific requirements of ML projects.
Key practices that are well established in traditional software development, such as those related to \textsc{DevOps}~\citep{leite2019survey}, have been adapted to the ML domain (e.g., \textsc{MLOps}~\citep{makinen2021needs}).
In particular,Continuous Integration (CI) practices have become pivotal for enhancing the development efficiency and release stability of ML projects\citep{gift2021practical}.

CI is a widely adopted practice in software development that involves frequent and automated integration of code changes into a shared repository, at least daily~\citep{fowler-ci-2006, duvall2007continuous}. 
In the ML domain, \cite{rzig2022characterizing} conducted an analysis revealing that approximately 37\% of ML projects have integrated a CI service into their workflows. The fundamental philosophy behind CI is to ensure that the software passes all tests and can build at all times \citep{duvall2007continuous}.
By fostering frequent commits, shorter build durations, and high test coverage, CI enhances development workflows and project reliability \citep{soares2022effects}. However, while extensively studied in traditional software projects \citep{vasilescu2015quality, hilton2016usage, zhao2017impact, bernardo2018studying, nery2019empirical, santos2022investigating, saraiva2023unveiling}, the integration of CI into ML projects remains relatively unexplored.

Unlike traditional CI for non-ML projects, CI in ML environments needs to handle the inherently probabilistic nature of ML components~\citep{renggli2019continuous}.
As such, in the context of ML, CI extends beyond merely testing and validating code and components and also involves testing and validation of data, data schemas, and models~\citep{karamitsos2020applying}. 
Our previous work conducted a quantitative analysis of 93 ML projects and 92 non-ML projects, examining the differences in the adoption of four key CI practices (i.e., frequent code integration, short build duration, quick build fixes, and comprehensive test coverage)  between these two types of projects.
Our findings indicated that ML projects often require longer build durations, while some exhibit lower test coverage compared to non-ML projects~\citep{bernardo2024machine}.
While these insights are valuable, the underlying reasons for the observed differences in build duration remain unclear.
A deeper understanding of the specific barriers faced by ML projects in adopting CI is crucial to inform practitioners and enable the full potential of CI in improving code quality and overall project success.

The general goal of this study is to deepen the understanding of the perceptions of ML practitioners regarding the differences, specific barriers, and strategies associated with CI adoption in ML projects.
Through thematic analysis of survey responses from 155 ML practitioners, we identify key themes -- such as computational complexity, dependency management, and data handling -- that contribute to extended build durations and reduced test coverage in ML projects.
These findings provide actionable insights for improving CI in the ML domain, fostering more efficient CI workflows, and advancing the integration of CI practices in this critical area of software development.
By bridging existing quantitative findings with practitioners' experiences, this study offers valuable guidance for researchers and practitioners aiming to optimize CI workflows for ML projects.

Our investigation is guided by the following Research Questions (RQs):

\begin{itemize}
    \item \textit{\textbf{\RQone}}
    \newline
    \textbf{Motivation:} 
    Our prior work~\citep{bernardo2024machine} identified significant differences in CI practices between ML and non-ML projects. However, the reasons behind these differences remain unclear. Gaining insights from ML practitioners' perspectives can shed light on the unique challenges of CI adoption in ML projects and guide the development of tailored solutions.
    
    \textbf{Findings:} 
    Participants of our survey identified key differences in CI practices between ML and non-ML projects, particularly in test coverage (61.9\%) and build duration (63.2\%), while perceptions regarding fixing broken builds and commit frequency were more neutral. Challenges in CI implementation mentioned by the participants were categorized into eight main themes, including testing complexity, infrastructure requirements, and build duration stability. Co-occurrence analysis revealed strong interdependencies among these challenges, with resource and infrastructure requirements playing a central role in CI effectiveness. Our results emphasize the need for a holistic approach that considers testing, resource allocation, and model handling to enhance CI practices in ML projects.
    
    \item \textit{\textbf{\RQtwo}}
    \newline
    \textbf{Motivation:} 
    Build duration is a critical aspect of CI pipeline efficiency, directly affecting developer productivity and the pace of iteration cycles. While our previous study observed that ML projects often experience longer build times, the factors that ML practitioners associate with extended build durations remain unexplored.

    \textbf{Findings:}     
    Short build durations are generally considered important by 69\% of participants, though tolerance increases with project size and complexity. 75\% of participants expect ML projects to have longer build durations than non-ML projects, largely due to higher project complexity, model training demands, increased computational resource needs, extensive data handling, and dependency management.     
    
    \item \textit{\textbf{\RQthree}}
    \newline
    \textbf{Motivation:} 
    Test coverage is a crucial metric for ensuring software reliability. However, ML projects face lower test coverage rates compared to non-ML projects ~\citep{bernardo2024machine}. Therefore, This RQ aims to investigate whether test coverage is handled differently in ML projects, providing insights into the specific obstacles when testing ML projects.

    \textbf{Findings:} 
    Testing in ML projects is particularly challenging due to test complexity, data dependencies, the non-deterministic nature of ML systems, and computational resource constraints. Key challenges such as test reproducibility, data dependencies, and resource demands often overlap, highlighting the need for holistic testing approaches. While most practitioners (63\%) preferred high test coverage (70–100\%), a notable portion (20\%) accepted moderate coverage (50–70\%), citing ML-specific constraints.
\end{itemize}

\subsection*{\textbf{Paper organization}}

The remainder of this paper is organized as follows: 
Section \ref{sec:related_work} presents an overview of related work, highlighting key studies and prior research that establish the foundation and contextual motivation for our investigation.
Section \ref{sec:methodology} describes the study methodology, detailing the studied projects, datasets, and analytical approaches employed in our investigation. 
Section \ref{sec:demographics} presents the demographics of the study participants.
In Section \ref{sec:results}, we present the results of our analysis, highlighting key findings and patterns observed in the data. 
Section \ref{sec:discussion} delves into the implications of the results for research and practice. Finally, Section \ref{sec:limitations} examines the limitations of our study while Section \ref{sec:conclusion} concludes the paper.
\section{Background}
\label{sec:related_work}

In this section, we introduce the background and definitions related to the concepts of CI and MLOps and position our study within the context of prior research.

\subsection{MLOps in a nutshell}

The primary goal of industrial ML projects is to develop and deploy ML-based products rapidly into production environments~\citep{kreuzberger2023machine}. 
However, incorporating ML models into production remains a significant challenge~\citep{symeonidis2022mlops}. 
Similar to traditional software projects, where the development and operation teams often struggle with collaboration and software release updates, ML projects face additional hurdles, particularly in deploying models to production~\citep{symeonidis2022mlops}.
While ML projects share core characteristics with traditional software projects, they differ fundamentally due to their reliance on ML models as key components~\citep{gift2021practical}, introducing complexities such as model validation into the integration and deployment process.

To address these challenges, ML engineers, data scientists, front-end developers, and production engineers have been collaborating to streamline the deployment of ML projects, leading in the emergence of MLOps~\citep{symeonidis2022mlops}.
MLOps, short for Machine Learning Operations, refers to a set of practices and tools designed to facilitate the deployment and maintenance of ML projects in production environments~\citep{makinen2021needs}. 
MLOps extends DevOps principles by incorporating additional processes specific to ML, such as data validation, model training, and model monitoring~\citep{symeonidis2022mlops, makinen2021needs, calefato2022preliminary}.

DevOps is a methodology that fosters collaboration, communication, and integration between development (Dev) and operations (Ops) teams, aiming to streamline the process of delivering software~\citep{ebert2016devops}.
According to \cite{leite2019survey}, ``DevOps is a collaborative and multidisciplinary organizational effort to automate the continuous delivery of new software updates while guaranteeing their correctness and reliability''.
MLOps builds on this philosophy by advocating for automation and monitoring all stages of the ML project life cycle, including integration, testing, release, deployment, and infrastructure management~\citep{karamitsos2020applying}. 
Implementing an MLOps pipeline is particularly beneficial for organizations transitioning from ML proof-of-concept implementations to production-ready systems~\citep{makinen2021needs}.


\subsubsection{Continuous Integration in MLOps}

CI is a core principle of DevOps, aimed at ensuring that code changes are frequently integrated, tested, and validated in a shared repository~\citep{fowler-ci-2006, duvall2007continuous}. 
Originated from the agile XP methodology, CI requires that developers integrate new code into a shared repository, at least daily~\citep{fowler-ci-2006, duvall2007continuous}. 
According to \cite{beck2000extreme}, ``New code is integrated with the current system after no more than a few hours. When integrating, the whole system is built from scratch and all tests must pass or the changes are discarded". 
The fundamental philosophy behind CI is to ensure that the shared state of the software's codebase remains in a working state at all times~\citep{duvall2007continuous}. 

A well-implemented CI pipeline automatically compiles, tests, and packages software whenever changes are made~\citep{bernardo2023}. By enabling frequent, reliable updates, CI helps software projects achieve shorter release cycles while maintaining high quality~\citep{bosch2014continuous}. As a foundational DevOps practice, CI plays a crucial role in enabling Continuous Delivery (CDE) and Continuous Deployment (CD)\citep{yarlagadda2018understanding, lwakatare2016relationship}. CDE extends CI by automating software release processes, while CD further automates deployment, ensuring that validated updates reach end users seamlessly\citep{karvonen2017systematic}.

In the ML domain, CI encompasses not only traditional software testing but also additional layers of validation, including data verification, schema validation, and model performance tracking~\citep{karamitsos2020applying}. Given that MLOps is built on DevOps principles, integrating CI into ML pipelines is crucial for ensuring the reliability and reproducibility of ML workflows.


\subsubsection{Continuous Integration practices}

CI extends beyond the use of tools to automate build and testing processes, encompassing broader values and principles that shape its effective implementation.
\cite{duvall2007continuous} outlined seven best practices for teams implementing CI in their projects: (i) commit code frequently, (ii) avoid committing broken code, (iii) fix broken builds immediately, (iv) write automated developer tests, (v) ensure that all tests and inspections pass, (vi) run private builds, and (vii) prevent the propagation of broken code.

A literature review by \cite{staahl2014modeling} examined the varying interpretations and implementations of CI practices. The study found that CI adoption is highly context-dependent, with differences in test frequency, integration flows, and overall practices based on project requirements. Despite established guidelines, there is no universal consensus on the exact CI metrics projects should follow, such as the optimal number of daily commits or the ideal test execution strategy. These variations highlight the challenges in defining standardized CI practices~\citep{santos2024needMonitorCI}.

In the ML domain, recent studies propose tailored CI practices to address the unique challenges of ML workflows. For instance, Fowler introduces the concept of Continuous Delivery for Machine Learning (\textsc{CD4ML}), emphasizing reproducible model training, validation data management, and model quality assurance~\citep{Fowler2019CD4ML}. Additionally, \cite{bagai2024implementing} explore CI/CD strategies for ML in cloud environments, highlighting best practices such as automated testing and validation, infrastructure as code, version control, and containerization. 
Additionally, \cite{garg2021continuous} emphasize the importance of model monitoring in ML CI pipelines to detect model performance degradation.
These practices recommendations underscore the need for specialized approaches to CI in ML, ensuring robustness and reliability in dynamic and data and model-driven environments.


\subsection{Related Work}

The adoption of CI has been extensively studied in the context of traditional open-source projects, demonstrating its positive impact on software development practices. 
Previous research has highlighted the benefits of adopting a CI service in the development cycle of software projects, including improved bug detection~\citep{vasilescu2015quality}, increased release frequency~\citep{hilton2016usage}, higher throughput of PRs delivered per release~\citep{bernardo2018studying}, and better test coverage rates~\citep{saraiva2023unveiling}. 

Beyond investigating the adoption of CI services, recent studies have focused on specific CI practices employed by software projects~\citep{felidre2019continuous, santos2022investigating}. For instance, \cite{felidre2019continuous} examined 1,270 open-source projects using \textsc{Travis CI} to identify unhealthy CI practices, such as infrequent commits, lengthy build durations, and poor test coverage. They found that, in most projects, builds were completed under the 10-minute rule of thumb. In addition, \cite{santos2022investigating} conducted a quantitative analysis of CI practices by examining data from 90 open-source projects over two years, exploring the relationship between these practices and project productivity and quality.

In the ML domain, recent studies have focused on defining \textsc{MLOps}, exploring its tools, architectures, and associated challenges~\citep{alla2021mlops, symeonidis2022mlops, kreuzberger2023machine}. \cite{karamitsos2020applying} proposed practical techniques for integrating DevOps principles and CI/CD practices into ML applications, addressing the unique requirements of ML workflows. Similarly, \cite{calefato2022preliminary} examined the role of \textsc{MLOps} in automating critical tasks involved in building and deploying ML-enabled systems, highlighting the importance of automation in managing the complexity of ML pipelines. \cite{makinen2021needs} further underscored the significance of \textsc{MLOps} in data science, presenting findings from a global survey of 331 professionals across 63 countries.

\cite{rzig2022characterizing} conducted a large-scale analysis of 4,031 ML projects hosted on \textsc{GitHub}, revealing that only 37\% of ML projects had adopted CI services. This relatively low adoption rate highlights the challenges posed by the unique characteristics of ML workflows. Despite the availability of popular CI services like \textsc{GitHub Actions}, recent efforts have focused on developing dedicated CI solutions tailored to the specific needs of ML projects. For example, \cite{renggli2019continuous} introduced \textsc{Ease.ML/CI}, a tool designed to prevent overfitting during iterative ML development. Likewise, \cite{karlavs2020building} proposed specialized CI services that address the probabilistic and iterative nature of ML workflows.
In addition, in our prior work~\citep{bernardo2024machine}, we provided a comparative quantitative analysis of CI practices in ML and non-ML projects, identifying longer build durations and lower test coverage rates in ML projects, particularly in medium-sized ones. 
These findings suggest that the adoption of CI in ML projects might be challenged by unique technical and workflow constraints that are not present in traditional software development.

While existing research has primarily provided quantitative assessments of CI adoption trends or proposed tools to address specific ML-related challenges, our study takes a qualitative perspective, focusing on ML practitioners' experiences with CI adoption. Specifically, we explore the underlying reasons behind the differences in CI adoption between ML and non-ML projects, the challenges ML projects face in building and testing their components, and the strategies practitioners employ to integrate CI practices effectively into ML projects. By capturing practitioner insights, our study bridges the gap between previous quantitative observations and real-world implementation challenges, offering a deeper understanding of the practical barriers to CI adoption in ML projects.


\section{Research Methodology}
\label{sec:methodology}

In this section, we describe how we selected the ML practitioners for our study, the data collection process, and the research approach we used to perform our analysis.

\subsection{Subject Projects}

Our prior work~\citep{bernardo2024machine} quantitatively analyzed the adoption of CI practices across 93 ML projects and 92 non-ML projects. This study revealed that ML projects often experience longer build durations and that medium-sized ML projects tend to have lower test coverage compared to non-ML projects. Building on these findings, our current study seeks to deepen our understanding of the specific factors driving these differences in CI practice adoption. To achieve this, we employed a qualitative, survey-based approach, focusing on the perceptions of ML practitioners (i.e., contributors and integrators of ML projects) regarding the challenges, barriers, and strategies associated with CI adoption in ML projects.

To maintain consistency with prior analyses and minimize potential biases associated with using an unverified or outdated collection of projects, we based our study on the 
93 ML projects investigated in our prior work~\citep{bernardo2024machine}. 
This dataset is both up-to-date and meticulously curated, representing a diverse collection of actively maintained ML projects that successfully integrate CI workflows into their pipelines. Additionally, the projects are categorized by size—small, medium, or large—based on their Lines of Code (LOC), ensuring a comprehensive representation across varying scales of ML development.

In addition to investigating general differences in CI adoption between ML and non-ML projects, this study focuses on the factors influencing build durations in ML projects, as outlined in RQ2. 
To capture diverse and meaningful insights, we adopted a sampling approach designed to reflect a wide range of experiences related to build durations. Specifically, we targeted projects with the shortest and those with the longest build durations to ensure that our analysis encompasses the full spectrum of challenges and characteristics encountered in ML workflows.

To sample practitioners for the survey, we selected those that are associated with the top 25\% of projects with the shortest build durations and the top 25\% with the longest build durations. 
As a result, we identified practitioners from 47 ML projects.
This targeted selection enhances the relevance of our findings by ensuring representation from both extremes of build duration characteristics, providing a comprehensive view of the characteristics and challenges faced in the CI workflows of ML projects. The list of the investigated ML projects, along with their characteristics (e.g., size and median build duration), is provided in Table~\ref{tab:project_characteristics}. 

\begin{table}[H]
  \centering
  \caption{Characteristics of Investigated Machine Learning Projects.}
    \begin{tabular}{cp{3.6cm}
    >{\raggedright\arraybackslash}p{2cm}
    >{\raggedleft\arraybackslash}p{2cm}
    >{\raggedright\arraybackslash}p{2cm}}
    \toprule
    \multirow[t]{3}{*}{\textbf{\#}} & \multirow[t]{3}{*}{\textbf{Project}} & \multirow[t]{3}{*}{\textbf{LOC Size}} & \textbf{Median build duration (minutes)} & \textbf{Build duration category} \bigstrut\\
    \midrule
    \textbf{1} & alan-turing-institute/sktime & large & 127.3 & longer \bigstrut[t]\\
    \textbf{2} & amark/gun & large & 1.3   & shorter \\
    \textbf{3} & apache/incubator-mxnet & large & 192.8 & longer \\
    \textbf{4} & apache/spark & large & 123.7 & longer \\
    \textbf{5} & apache/superset & large & 10.2  & shorter \\
    \textbf{6} & AUTOMATIC1111/stable-diffusion-webui & medium & 7.4   & shorter \\
    \textbf{7} & BehaviorTree/BehaviorTree.CPP & medium & 3.7   & shorter \\
    \textbf{8} & BLKSerene/Wordless & medium & 37.4  & longer \\
    \textbf{9} & chakki-works/doccano & medium & 2.8   & shorter \\
    \textbf{10} & criteo/tf-yarn & small & 2.4   & shorter \\
    \textbf{11} & DandyDev/slack-machine & small & 2.5   & shorter \\
    \textbf{12} & diffgram/diffgram & large & 5.1   & shorter \\
    \textbf{13} & dmlc/tvm & large & 90.2  & longer \\
    \textbf{14} & FluxML/Metalhead.jl & small & 28.4  & longer \\
    \textbf{15} & FluxML/NNlib.jl & medium & 37.4  & longer \\
    \textbf{16} & huggingface/pytorch-pretrained-BERT & large & 5.4   & shorter \\
    \textbf{17} & huggingface/transformers & large & 5.4   & shorter \\
    \textbf{18} & JohnSnowLabs/spark-nlp & large & 47.0    & longer \\
    \textbf{19} & jtablesaw/tablesaw & medium & 4.3   & shorter \\
    \textbf{20} & kendryte/nncase & large & 147.4 & longer \\
    \textbf{21} & LaurentMazare/tch-rs & large & 7.1   & shorter \\
    \textbf{22} & microsoft/dowhy & medium & 29.8  & longer \\
    \textbf{23} & microsoft/LightGBM & medium & 21.8  & longer \\
    \textbf{24} & microsoft/onnxruntime & large & 99.6  & longer \\
    \textbf{25} & microsoft/pai & medium & 3.9   & shorter \\
    \textbf{26} & mlpack/mlpack & large & 126.1 & longer \\
    \textbf{27} & mne-tools/mne-cpp & large & 62.4  & longer \\
    \textbf{28} & msdslab/automated-systematic-review & medium & 4.7   & shorter \\
    \textbf{29} & nilearn/nilearn & medium & 31.0    & longer \\
    \textbf{30} & opencv/dldt & large & 2.5   & shorter \\
    \textbf{31} & OpenKore/openkore & large & 5.0     & shorter \\
    \textbf{32} & OpenNMT/OpenNMT-py & medium & 3.4   & shorter \\
    \textbf{33} & pytorch/ignite & medium & 29.0    & longer \\
    \textbf{34} & pytorch/tnt & medium & 4.1   & shorter \\
    \textbf{35} & RubixML/RubixML & medium & 3.6   & shorter \\
    \textbf{36} & scikit-learn/scikit-learn & large & 0.3   & shorter \\
    \textbf{37} & SeldonIO/seldon-core & large & 4.6   & shorter \\
    \textbf{38} & shimat/opencvsharp & medium & 20.0    & longer \\
    \textbf{39} & skorch-dev/skorch & medium & 5.3   & shorter \\
    \textbf{40} & smistad/FAST & medium & 46.6  & longer \\
    \textbf{41} & sorgerlab/indra & medium & 19.3  & longer \\
    \textbf{42} & tensorflow/addons & medium & 19.4  & longer \\
    \textbf{43} & tensorly/tensorly & medium & 46.6  & longer \\
    \textbf{44} & tesseract-ocr/tesseract & large & 124.9 & longer \\
    \textbf{45} & Texera/texera & large & 6.8   & shorter \\
    \textbf{46} & TuringLang/Turing.jl & small & 107.1 & longer \\
    \textbf{47} & zhenghaoz/gorse & medium & 6.9   & shorter \bigstrut[b]\\
    \bottomrule
    \end{tabular}%
  \label{tab:project_characteristics}%
\end{table}%

\subsection{Data Collection}

To identify practitioners within the 47 investigated ML projects, we focused on individuals who actively contributed to the projects after the adoption of \textsc{GitHub Actions} CI workflows. Specifically, we selected integrators who either merged or closed at least one pull request (PR) or submitted at least one PR that was successfully merged into the \textit{main/master} branch of the project codebase during this period. This approach ensures that the selected practitioners directly contributed to the project while CI workflows were in use, enabling them to provide relevant and informed insights into CI practices in ML projects.

We collected PR metadata for the studied projects using the \textsc{GitHub API} on June 14, 2024. 
The PR's metadata include details of the PR number, state, author login, base branch, number of additions, deletions, changed files, commit count, and whether the PR was merged or closed, along with the login of the user who closed it.
By analyzing data from the period following the adoption of \textsc{GitHub Actions}, we identified 114,598 PRs reviewed by 3,276 unique integrators. Additionally, 6,861 contributors had at least one PR successfully merged. Importantly, these represent two distinct groups of practitioners in our study: contributors, who authored PRs, and integrators, who merged or rejected them. Among the contributors, 1,909 also acted as integrators, reflecting some overlap between the two roles. After accounting for this overlap, we identified a total of 4,952 unique contributors. Combined with the 3,276 unique integrators, this results in 8,228 practitioners involved in the studied projects.

To contact the practitioners involved in these projects, we collected their email addresses using the \textsc{GitHub API}, ensuring we only collected publicly available information. Of the 3,276 unique integrators, we retrieved 2,060 email addresses; for the 4,952 contributors, we retrieved 2,947. In total, we collected 5,007 unique email addresses from practitioners in the 47 analyzed projects.

To collect our data, we designed a web-based survey and sent invitations by email to all 5,007 ML practitioners whose email addresses were available.
The invitation letter is included in Appendix~\ref{sec:appendix_invitation_email_example}. To encourage participation, we offered respondents the opportunity to win one of ten \$50 Amazon or Steam gift cards, distributed through a random drawing. Participants were eligible for the draw only if they completed all survey questions and explicitly indicated their willingness to participate.
We used \textsc{Mailgun}\footnote{\url{https://www.mailgun.com}} to send personalized email invitations to each practitioner.
If a practitioner was associated with multiple investigated projects, we sent only one invitation, prioritizing the project where they had the highest number of integrated PRs.

In total, we received 155 responses, resulting in a response rate of 3.1\% (\nicefrac{155}{5007}). These responses came from practitioners associated with 30 of the 47 investigated projects. Table~\ref{tab:number_responses_per_respondent_type} presents details on the number of practitioners contacted per project, the responses received, and the corresponding response rates. To maintain anonymity, practitioners' names have been replaced with unique identifiers in the table. For example, practitioner 01 is labeled as ``P1'' and is associated with the \textit{alan-turing-institute/sktime} project.

\begin{table}[H]
  \centering
  \caption{Number of responses and practitioners of the studied ML projects that were invited to participate.}
\begin{tabular}{cp{3cm}>{\raggedleft\arraybackslash}p{1.7cm}
>{\raggedleft\arraybackslash}p{1.5cm}
>{\raggedleft\arraybackslash}p{1.3cm}
>{\raggedright\arraybackslash}p{1.5cm}}
\toprule
\textbf{\#} & \textbf{Project} & \textbf{Number of practitioners} & \textbf{Number of responses} & {\textbf{Response rate}} & \textbf{Practitioner IDs} \\
\midrule
1 & alan-turing-institute/sktime & 123 & 8 & 6.5\% & P1--P8 \\
2 & amark/gun & 34 & 2 & 5.9\% & P9--P10 \\
3 & apache/spark & 189 & 1 & 0.5\% & P11--P11 \\
4 & apache/superset & 435 & 12 & 2.8\% & P12--P23 \\
5 & AUTOMATIC1111/stable-diffusion-webui & 293 & 8 & 2.7\% & P24--P31 \\
6 & BehaviorTree/BehaviorTree.CPP & 56 & 3 & 5.4\% & P32--P34 \\
7 & chakki-works/doccano & 37 & 1 & 2.7\% & P35--P35 \\
8 & diffgram/diffgram & 8 & 2 & 25.0\% & P36--P37 \\
9 & dmlc/tvm & 399 & 7 & 1.8\% & P38--P44 \\
10 & FluxML/NNlib.jl & 24 & 1 & 4.2\% & P45--P45 \\
11 & huggingface/pytorch-pretrained-BERT & 700 & 18 & 2.6\% & P46--P63 \\
12 & huggingface/transformers & 689 & 19 & 2.8\% & P64--P82 \\
13 & kendryte/nncase & 7 & 1 & 14.3\% & P83--P83 \\
14 & LaurentMazare/tch-rs & 39 & 2 & 5.1\% & P84--P85 \\
15 & microsoft/LightGBM & 89 & 2 & 2.2\% & P86--P87 \\
16 & microsoft/onnxruntime & 196 & 10 & 5.1\% & P88--P97 \\
17 & microsoft/pai & 12 & 1 & 8.3\% & P98--P98 \\
18 & mlpack/mlpack & 38 & 4 & 10.5\% & P99--P102 \\
19 & msdslab/automated-systematic-review & 23 & 1 & 4.3\% & P103--P103 \\
20 & nilearn/nilearn & 58 & 6 & 10.3\% & P104--P109 \\
21 & opencv/dldt & 394 & 4 & 1.0\% & P110--P113 \\
22 & pytorch/tnt & 10 & 1 & 10.0\% & P114--P114 \\
23 & scikit-learn/scikit-learn & 627 & 25 & 4.0\% & P115--P139 \\
24 & SeldonIO/seldon-core & 69 & 2 & 2.9\% & P140--P141 \\
25 & shimat/opencvsharp & 19 & 1 & 5.3\% & P142--P142 \\
26 & sorgerlab/indra & 9 & 1 & 11.1\% & P143--P143 \\
27 & tensorflow/addons & 50 & 3 & 6.0\% & P144--P146 \\
28 & tesseract-ocr/tesseract & 46 & 6 & 13.0\% & P147--P152 \\
29 & TuringLang/Turing.jl & 26 & 2 & 7.7\% & P153--P154 \\
30 & zhenghaoz/gorse & 24 & 1 & 4.2\% & P155--P155 \\
31 & apache/incubator-mxnet & 40 & 0 & 0.0\% & --- \\
32 & BLKSerene/Wordless & 3 & 0 & 0.0\% & --- \\
33 & criteo/tf-yarn & 4 & 0 & 0.0\% & --- \\
34 & DandyDev/slack-machine & 4 & 0 & 0.0\% & --- \\
35 & FluxML/Metalhead.jl & 12 & 0 & 0.0\% & --- \\
36 & JohnSnowLabs/spark-nlp & 30 & 0 & 0.0\% & --- \\
37 & jtablesaw/tablesaw & 18 & 0 & 0.0\% & --- \\
38 & microsoft/dowhy & 20 & 0 & 0.0\% & --- \\
39 & mne-tools/mne-cpp & 5 & 0 & 0.0\% & --- \\
40 & OpenKore/openkore & 11 & 0 & 0.0\% & --- \\
41 & OpenNMT/OpenNMT-py & 18 & 0 & 0.0\% & --- \\
42 & pytorch/ignite & 55 & 0 & 0.0\% & --- \\
43 & RubixML/RubixML & 11 & 0 & 0.0\% & --- \\
44 & skorch-dev/skorch & 9 & 0 & 0.0\% & --- \\
45 & smistad/FAST & 4 & 0 & 0.0\% & --- \\
46 & tensorly/tensorly & 16 & 0 & 0.0\% & --- \\
47 & Texera/texera & 24 & 0 & 0.0\% & --- \\
\midrule
& \textbf{Total} & \textbf{5,007} & \textbf{155} & \textbf{3.1\%} & \\
\bottomrule
\end{tabular}  \label{tab:number_responses_per_respondent_type}%
\end{table}%

Our survey is organized into five major sections, as described in Table \ref{tab:survey_structure}. It includes 20 questions, combining 8 closed- and 12 open-ended questions, designed to collect both quantitative and qualitative data. The estimated completion time is approximately 10 minutes.
To ensure relevance and foster more thoughtful responses, we designed 47 unique questionnaires, each tailored to specific characteristics and statistics of an associated project.
For example, in \textsc{Question \#4.3} of the form sent to the practitioners of the project \textit{tesseract-ocr/tesseract}, we asked: \textit{"When analyzing the data of the \textit{tesseract-ocr/tesseract} project, we observed that this project has a median build duration of 124.9 minutes, which is longer than 90\% of the investigated projects of similar size. Do you have any insights into why this project has a longer build duration?"}.
This customization allowed us to provide participants with context-specific data, enabling them to offer richer and more meaningful insights about their respective projects.
A complete example of the survey is available in our online Appendix\footnote{\url{https://zenodo.org/records/14902811}}, which includes the customized questionnaire sent to participants of the \textit{tesseract-ocr/tesseract}\footnote{\url{http://github.com/tesseract-ocr/tesseract}} project.

\begin{table}
\centering
\caption{Survey Structure and Description.}
\label{tab:survey_structure}
\begin{tabular}{p{3cm}p{8cm}}
\toprule
\textbf{Section} & \textbf{Description} \\
\midrule
Participant Information & Collects demographic data and information about participants' experience, including their experience contributing to ML projects, as well as familiarity with CI practices. \\
\midrule
Perceptions about CI Practices & Gathers insights into the challenges and differences when adopting CI practices in ML projects compared to non-ML projects. Focuses on team practices such as maintaining short build durations, frequent commits, and upholding high test coverage. \\
\midrule
Reflection on Previous Findings & Explores participants' views on results from our prior study, focusing on disparities in CI adoption between ML and non-ML projects. The questions explore the underlying factors contributing to differences in build durations, test coverage, and potential strategies for enhancement. \\
\midrule
Project-Specific Analysis & Presents data derived from a specific project (e.g., \textit{tesseract-ocr/tesseract}) to solicit feedback on unique challenges and techniques for enhancing CI practices in a real-world scenario. \\
\midrule
Conclusion and Follow-Up & Allows participants to opt into follow-up interviews, request updates on study findings, and share additional comments. Ensures eligibility for the gift card drawing by confirming survey completion. \\
\bottomrule
\end{tabular}
\end{table}

To encourage participation and a higher response rate, none of the questions in our survey were mandatory. As a result, the number of responses for each question varied, as not all participants answered every question. Responses were marked as ``NA'' (No Answer) if a participant left a question blank.
Table~\ref{tab:response_rates} provides a detailed overview of each survey question, including its description, type (open-ended or close-ended), and corresponding response rates. Close-ended questions generally achieved higher response rates, with several receiving complete responses (e.g., Questions 1.1–1.4 at 100\%). Conversely, open-ended questions exhibited slightly lower response rates, with the lowest being 85.2\% (Question 3.7). This trend indicates that open-ended questions, which typically require more effort and time to answer, may discourage some participants from responding. Nonetheless, the consistently high response rates across all question types highlight strong engagement from the participants.

\begin{table}
\centering
\caption{Survey Questions and Response Rates.}
\label{tab:response_rates}
\begin{tabular}{cp{6cm}lr}
\toprule
\textbf{\#} & \textbf{Question Description} & \textbf{Question Type} & \textbf{Responses (Rate)} \\
\midrule
1.1 & Experience developing software & Close-ended & 155/155 (100\%) \\
1.2 & Experience developing ML projects & Close-ended & 155/155 (100\%) \\
1.3 & Primary roles in ML projects & Close-ended & 155/155 (100\%) \\
1.4 & Familiarity with CI concepts & Close-ended & 155/155 (100\%) \\
2.1 & ML projects strive to incorporate CI practices & Open-ended & 146/155 (94.2\%) \\
2.2 & Challenges or differences when implementing a CI pipeline & Open-ended & 142/155 (91.6\%) \\
2.3 & ML projects commit more frequently & Close-ended & 149/155 (96.1\%) \\
2.4 & ML projects have longer build durations & Close-ended & 151/155 (97.4\%) \\
2.5 & ML projects have lower test coverage & Close-ended & 150/155 (96.8\%) \\
2.6 & ML projects fix broken builds more quickly & Close-ended & 149/155 (96.1\%) \\
3.1 & Importance of ML projects keeping a short build duration & Close-ended & 151/155 (97.4\%) \\
3.2 & Perceptions about previous study results on build duration in ML projects & Open-ended & 141/155 (91\%) \\
3.3 & Strategies to reduce build duration in ML projects & Open-ended & 138/155 (89\%) \\
3.4 & Acceptable test coverage rate for an ML project & Close-ended & 153/155 (98.7\%) \\
3.5 & Perceptions about previous study results on test coverage in ML projects & Open-ended & 135/155 (87.1\%) \\
3.6 & Challenges in testing ML projects & Open-ended & 133/155 (85.8\%) \\
3.7 & Strategies to enhance test coverage in ML projects & Open-ended & 132/155 (85.2\%) \\
4.1 & Familiarity with the CI pipeline of the studied project & Close-ended & 154/155 (99.4\%) \\
4.2 & Acceptable build duration for ML projects & Close-ended & 150/155 (96.8\%) \\
4.3 & Perceptions about the build duration of the studied project & Open-ended & 135/155 (87.1\%) \\
\bottomrule
\end{tabular}
\end{table}

\subsection{Analytical Approach}

We applied an inductive thematic analysis to identify, analyze, and report themes within the qualitative data collected from our questionnaire, following the approach outlined by \cite{braun2006using}. To ensure rigor and transparency in the process, we adhered to the guidelines proposed by \cite{nowell2017thematic}.

The initial step of our thematic analysis involved open-coding the qualitative data. This process refers to assigning codes to relevant segments of data collected from the responses to our open-ended survey questions. Each question was coded by at least two authors, enhancing the robustness of the analysis and mitigating potential bias.
The first author conducted open coding for all eight open-ended questions in the survey, and to ensure reliability in the coding process, the second and third authors coded responses for three questions, and the fourth author coded two questions. 
Afterwards, the fifth author reviewed the entire set of codes generated by the two coders of each question. This review process helped resolve disagreements or ambiguities, refine the coding, and add additional entries where necessary.

Once the coding process was completed, the first author performed axial coding, grouping codes into higher-level themes. These themes represented broader conceptual constructs, organizing multiple related codes under a common idea. For example, a single theme might encompass several related codes addressing a specific aspect of CI practices in ML projects.


Finally, we report the codes and themes derived from our thematic analysis in the results section.
When presenting our findings, we indicate the number of quotes associated with each code and theme using superscripts. 
However, it is important to note that these numbers do not necessarily indicate the relevance or significance of a code. For instance, a code may be cited in more quotes simply because it is more easily remembered by participants, rather than due to its importance.
To provide further context and depth, we include representative quotes from participants. To maintain anonymity, participant names are replaced with unique IDs.

\begin{figure}[H]
	\centering
	\includegraphics[width=0.7\textwidth]{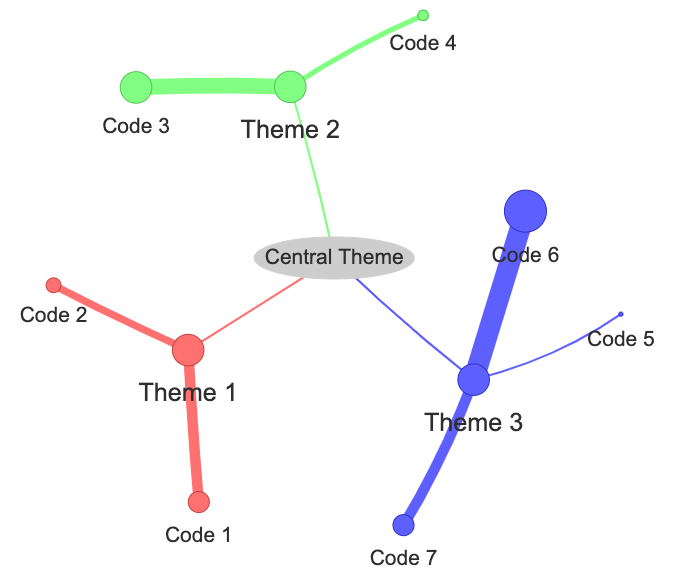}
	\caption{Example of a Network Mapping Chart visualizing the relationships between themes and codes in thematic analysis.}
	\label{fig:network_mapping_chart_example}       
\end{figure}

While textual representations highlight key insights, we also employ network mapping charts to provide a structured visual representation of the relationships between themes and codes.
At the center of the network lies the core theme, encapsulating the primary focus of the RQ. Surrounding it are second-level themes, which further break down into third-level themes (codes), organized based on their conceptual relationships. Figure~\ref{fig:network_mapping_chart_example} presents an example of a Network Mapping Chart, illustrating these relationships in the thematic analysis.
Each third-level theme (code) offers granular insights into specific aspects of the data. The thickness of the edges in the network represents frequency, indicating how prominently each code appeared during the analysis.

\section{Demographics of Study Participants}
\label{sec:demographics}

This section provides an overview of the demographics of our study participants, highlighting their general experience in software development, their experience in ML projects, their primary roles and activities in ML software development, their familiarity with CI concepts, and their knowledge of the CI pipelines employed in the ML projects they contribute to.

\begin{figure}
	\centering
	\includegraphics[width=12cm]{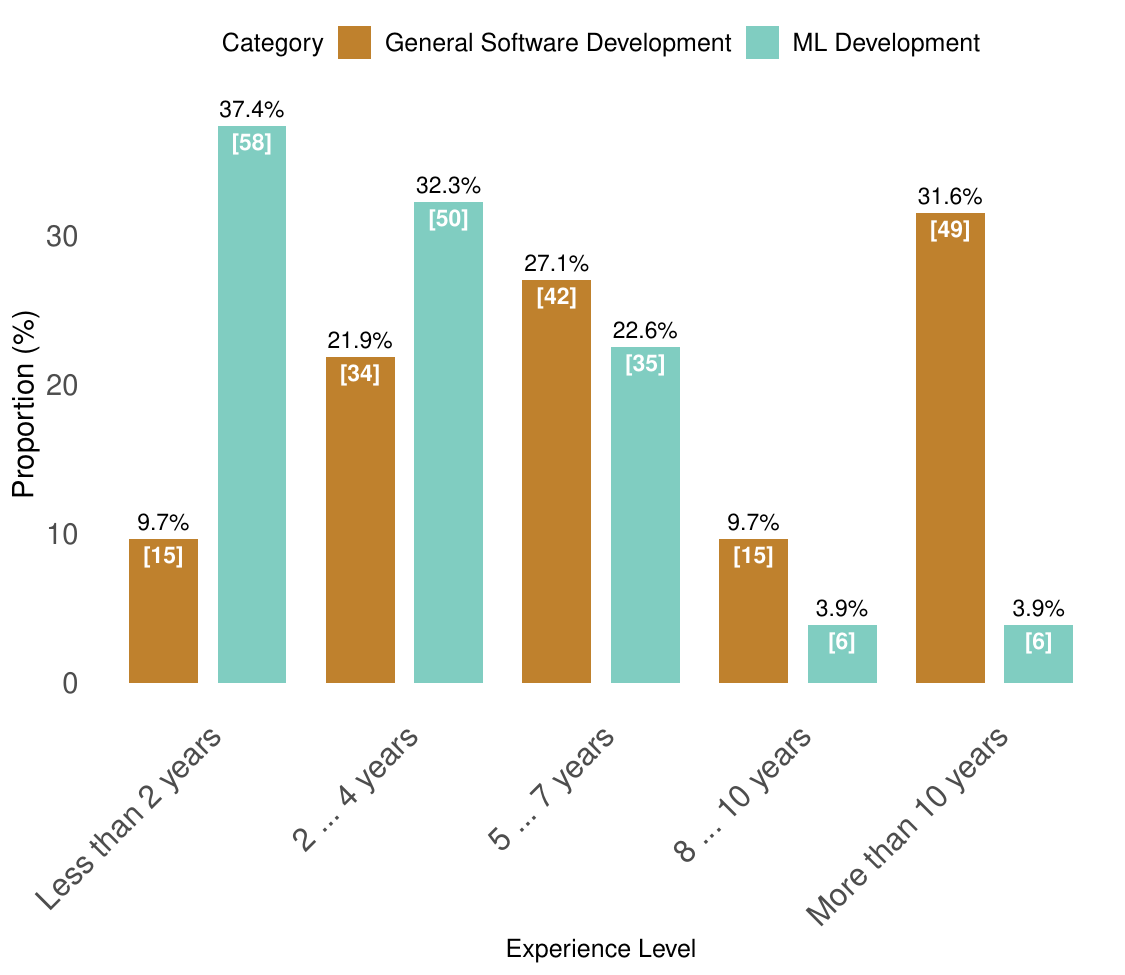}
	\caption{Participants' experience in general software development and ML project development.}
	\label{fig:q1_1_and_q1_2_developer_expericel_level}       
\end{figure}

Figure~\ref{fig:q1_1_and_q1_2_developer_expericel_level} presents the general software development experience of the study participants and their experience developing ML projects.
The results indicate that the majority of participants (68.4\%, \nicefrac{106}{155}) have five or more years of general software development experience, with 31.6\% of the participants having ten or more years of experience. This highlights the relatively high level of expertise of our participants, suggesting that the insights drawn from their responses are informed by substantial professional experience. Nevertheless, the collected responses encompass a variety of perspectives that also include those of less experienced developers.

Regarding experience in ML projects, the responses reveal a balanced distribution: 37.4\% of participants have less than 2 years of experience, 32.3\% have 2 to 4 years, and 30.4\% have 5 or more years of experience.
This distribution reflects the relative recency of ML projects as a field, where it is expected that many practitioners are still accumulating experience. While a significant portion of participants has considerable expertise, the presence of less experienced practitioners is equally valuable. These participants might provide unique perspectives on the challenges faced when starting to work on ML projects, particularly in adopting and using CI pipelines.


\begin{figure}
	\centering
	\includegraphics[width=12cm]{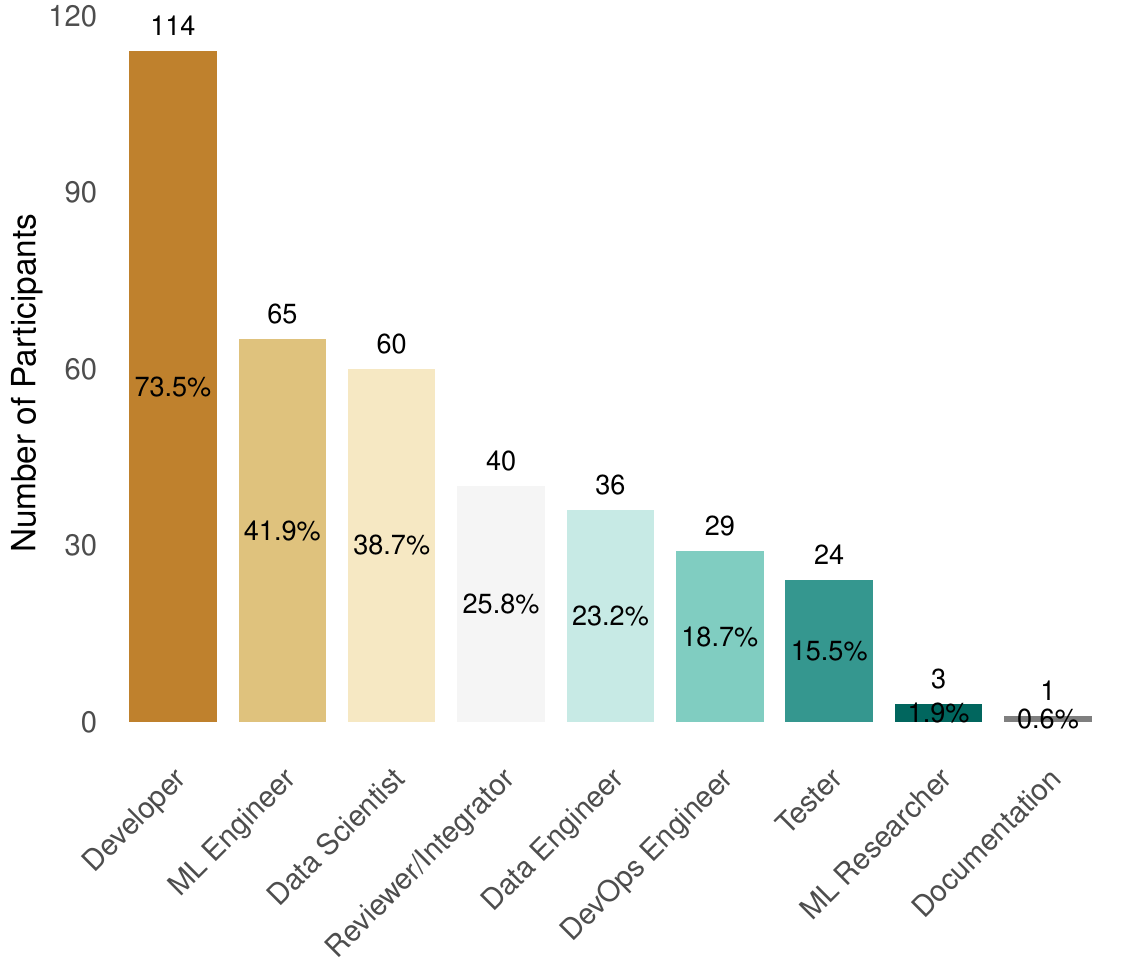}
	\caption{Participants' primary roles in ML projects.}
	\label{fig:q1_3_Primary_roles_in_ML_projects}       
\end{figure}

The distribution of participants' primary roles in ML projects, as shown in Figure~\ref{fig:q1_3_Primary_roles_in_ML_projects}, highlights the diverse expertise within our dataset.
Participants could select multiple roles in our form. Consequently, the reported percentages do not sum to 100\%, reflecting the fact that individuals often assume multiple responsibilities in ML projects.
The majority of participants identify as Developers (73\%, 114 participants), emphasizing their central role in ML project workflows. ML Engineers, responsible for model deployment and monitoring, comprise a significant portion of the participants (42\%, 65 individuals), while Data Scientists, focused on model development and validation, represent 39\% (60 participants).

Participants involved in code review and integration, crucial for maintaining code quality and consistency, account for 26\% (40 participants). Data Engineers, handling tasks such as data ingestion and storage, make up 23\% (36 participants). DevOps Engineers, who oversee the deployment and maintenance of CI/CD pipelines, constitute 19\% (29 participants). A smaller but essential group includes Testers (15\%, 24 participants), who ensure quality through systematic testing, and ML Researchers (1.9\%, 3 participants), who focus on advancing the theoretical aspects of ML. Finally, documentation roles, while less represented, remain critical for maintaining project records, with 0.6\% (1 participant) listing this as their primary responsibility.
This broad range of roles reflects the diverse expertise of our participants, capturing insights across the full lifecycle of ML project development and deployment.

\begin{figure}
	\centering
	\includegraphics[ width=12cm]{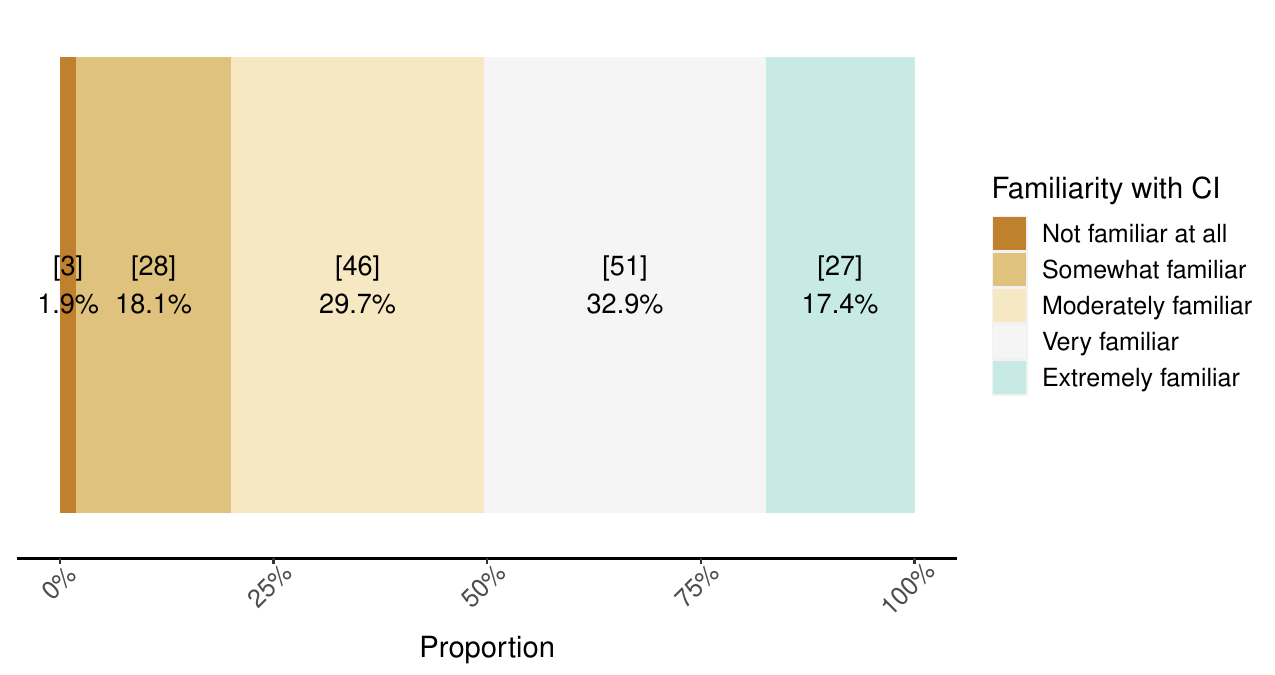}
	\caption{Participants' familiarity with CI.}
	\label{fig:q1_4_Familiarity_with_CI_concepts}       
\end{figure}

Regarding participants' familiarity with CI, we found that most have a strong understanding of the concept, which enhances the credibility and depth of their responses on CI practices in both ML and non-ML projects.
Figure~\ref{fig:q1_4_Familiarity_with_CI_concepts} illustrates the distribution of participants' familiarity levels with CI. A significant proportion reported a high level of familiarity, with 32.9\% (51 participants) indicating they are ``Very familiar'' and 17.4\% (27 participants) describing themselves as ``Extremely familiar''. Additionally, 29.7\% (46 participants) rated their familiarity as ``Moderately familiar''. Lower levels of familiarity were reported by smaller groups, with 18.1\% (28 participants) identifying as ``Somewhat familiar'' and only 1.9\% (3 participants, comprising 2 Developers and 1 Data Scientist primarily involved in model development and validation) as ``Not familiar at all''.

\begin{figure}
	\centering
	\includegraphics[width=12cm]{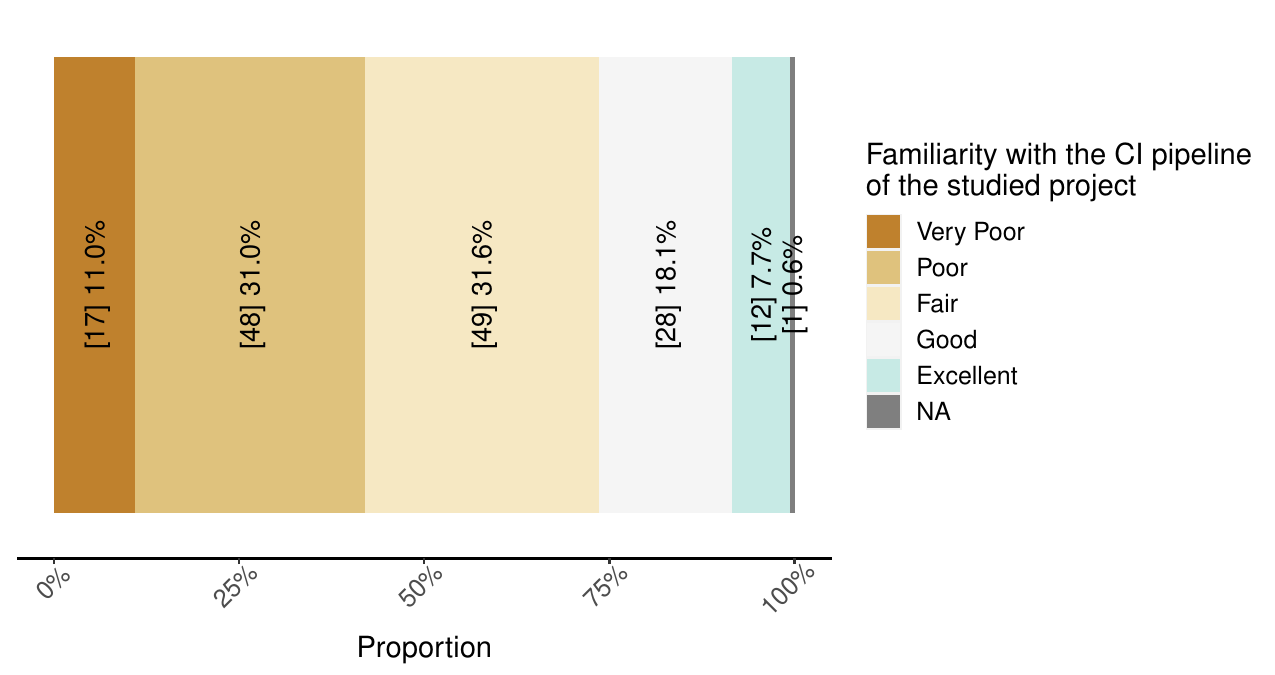}
	\caption{Participants' familiarity with the CI pipeline of the studied projects.}	\label{fig:q4_1_Familiarity_with_the_CI_pipeline_of_the_studied_project}       
\end{figure}

Beyond general familiarity with CI concepts, Figure~\ref{fig:q4_1_Familiarity_with_the_CI_pipeline_of_the_studied_project} reveals that most participants also possess a strong understanding of the CI pipelines used in the studied ML projects. Specifically, 31.6\% rated their familiarity as ``Fair'', 31.0\% as ``Good'', and 18.1\% as ``Excellent''. 
Only a smaller proportion of participants reported limited familiarity, with 11.0\% rating it as ``Poor'' and 7.7\% as ``Very Poor'', while an additional 7.7\% did not respond (NA). 
These findings highlight the strength of the dataset, as the participants' high level of familiarity ensures they are well-equipped to provide valuable and informed insights into the CI pipelines of ML projects, thereby enhancing the reliability and depth of their responses.

Additionally, in \textsc{Question \#2.1}, we asked participants whether they felt their teams were committed to key CI practices, including frequent commits, maintaining short build durations, promptly addressing broken builds, and ensuring high test coverage. The results show that 63.9\% (99 out of 155) of participants reported that their teams consistently prioritize these practices. In contrast, 11.6\% (18 out of 155) indicated partial commitment, with CI practices applied inconsistently or skipped in certain situations. Additionally, 14.2\% (22 out of 155) stated that their teams showed no commitment to CI practices, while 10.3\% (16 out of 155) did not provide a response.
These findings provide crucial context for understanding the broader landscape of CI adoption within ML project. Projects actively engaging with CI practices form a valuable foundation for participants’ responses, as they reflect both individual familiarity and team-level implementation of CI.

\section{\textbf{Results}}
\label{sec:results}

This section presents the findings for each research question addressed in the study.

\subsection*{\textbf{\RQone}}

\begin{figure}
	\centering
	\includegraphics[ width=12cm]{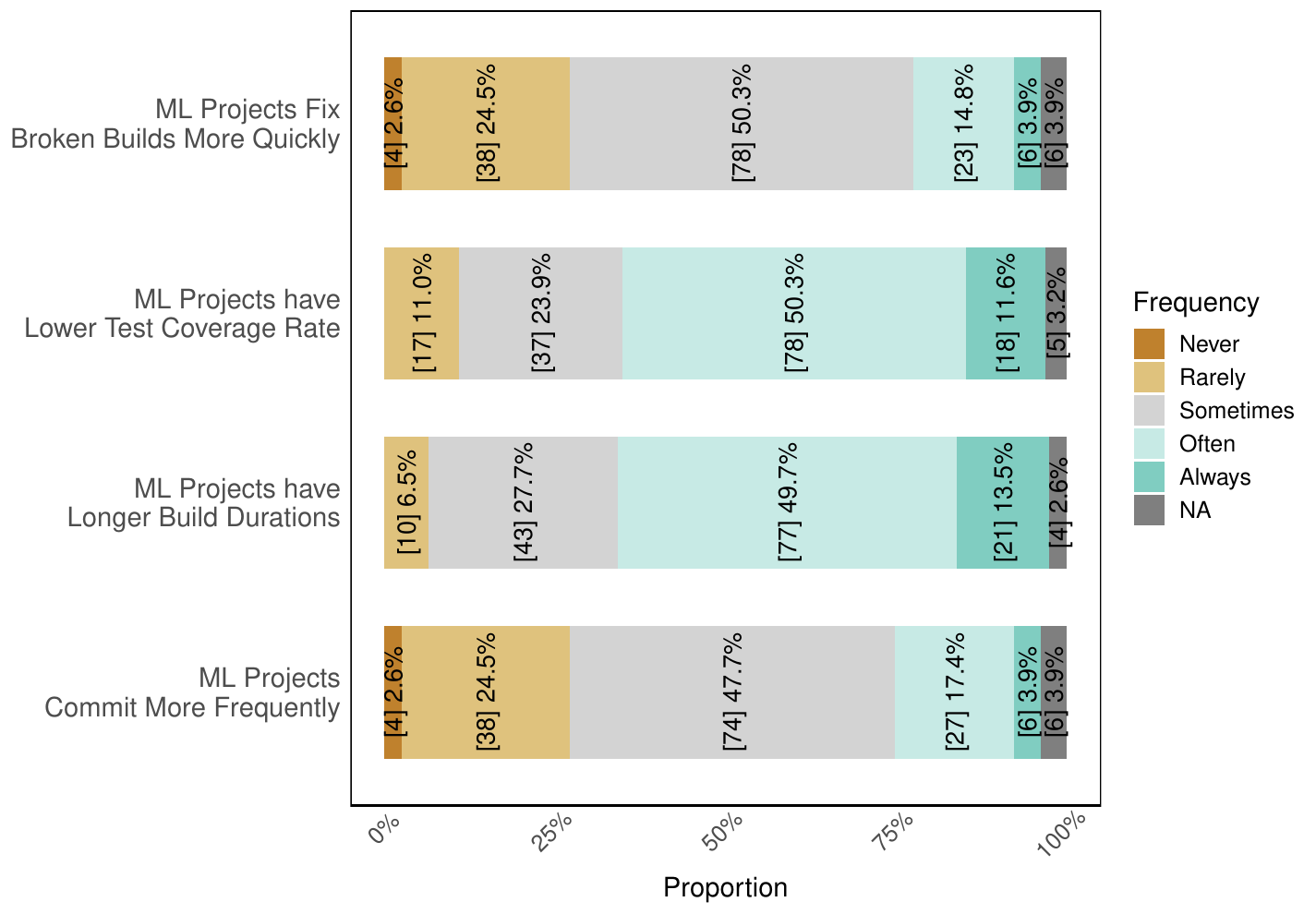}
	\caption{Participants' perception about differences in CI practices on ML projects.}
	\label{fig:q2_ci_practices_likert_scale}       
\end{figure}

In \textsc{Questions \#2.3 to \#2.6} of our survey, participants were asked how frequently they observe ML projects having more frequent commits, longer build durations, lower test coverage rates, and quicker fixes for broken builds compared to non-ML projects. As shown in Figure \ref{fig:q2_ci_practices_likert_scale}, the majority of participants often or always perceive ML projects as having longer build durations (63.2\%, \nicefrac{98}{155}) and lower test coverage (61.9\%, \nicefrac{96}{155}). 
In contrast, responses regarding commit frequency and the time to fix builds were more neutral. Most participants reported that ML projects only sometimes, rarely, or never fix broken builds more quickly (77.4\%, \nicefrac{120}{155}) or commit more frequently (74.8\%, \nicefrac{116}{155}) than non-ML projects.
Thus, the practices of fixing broken builds quickly and committing code frequently show a smaller perceived difference when compared to non-ML projects. These findings strongly align with our previous work \citep{bernardo2024machine}, which found statistically significant differences in build duration and test coverage between ML and non-ML projects, but no significant differences in the time taken to fix broken builds or in the commit frequency.

Another analysis for this RQ focuses on identifying unique challenges and differences when implementing a CI pipeline in ML projects compared to non-ML projects. In \textsc{Question \#2.2}, participants were asked to describe the perceived challenges and differences they experienced when adopting CI in ML projects, which we subsequently analyzed through a thematic analysis.


\begin{figure}
	\centering
	\includegraphics[ width=12cm]{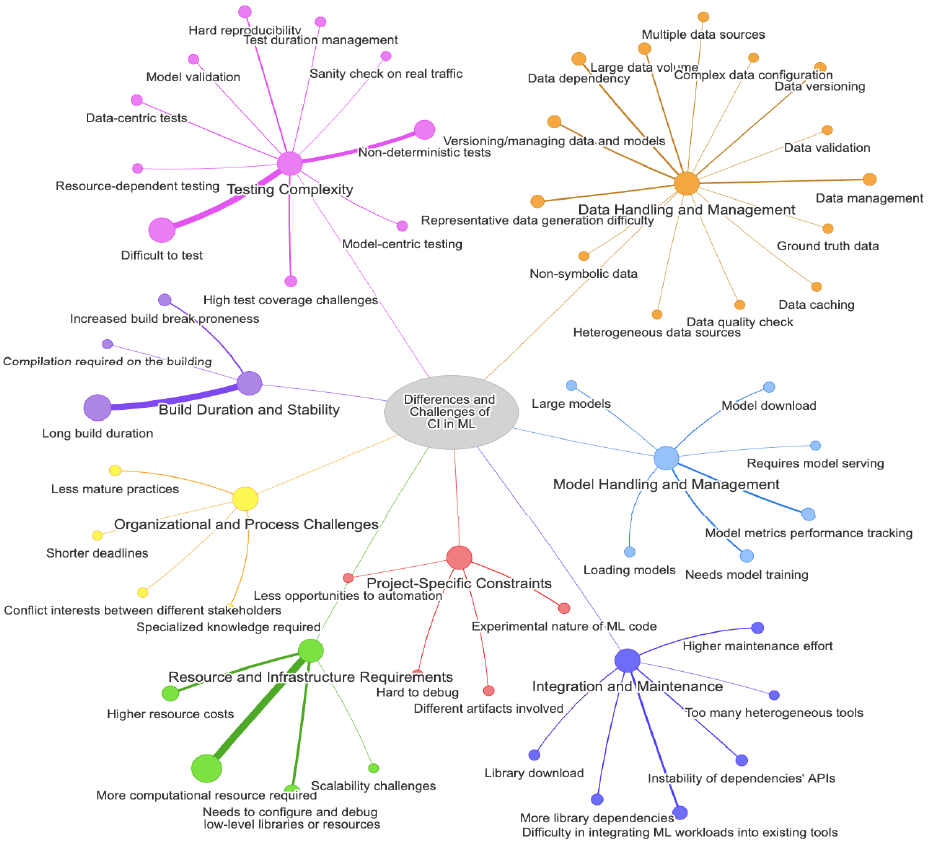}
	\caption{Perceived differences and challenges when adopting a CI pipeline in ML projects.}
	\label{fig:q2_2_differences_and_challenges}       
\end{figure}

Figure \ref{fig:q2_2_differences_and_challenges} illustrates the 8 themes and 51 codes that emerged from our analysis, highlighting key differences and challenges in the CI pipeline of ML projects. The themes are related to
\textit{Testing Complexity},
\textit{Infrastructure Requirements},
\textit{Build Duration and Stability},
\textit{Data Handling and Management},
\textit{Model Handling and Management},
\textit{Integration and Maintenance Challenges},
\textit{Organizational and Process Challenges}, and
\textit{Project-Specific Constraints}.
Additionally, 11 (7.1\%) participants perceived no significant differences in the challenges between the CI pipeline of ML and non-ML projects, while 28 (18.1\%) participants did not respond to this question. 
Below, we provide detailed insights into the main codes associated with each theme identified in our thematic analysis.

\textbf{\textit{Testing Complexity.}}\textsuperscript{(44)}
This is the most mentioned theme by our participants, which is related to the difficulties of ML projects in establishing robust testing within the CI pipeline. Participants frequently cited \textit{difficulties in testing}\textsuperscript{(26)}, \textit{non-determinism}, \textsuperscript{(17)} and \textit{hard reproducibility}\textsuperscript{(5)} as major hurdles. This theme reinforces the idea that \textit{achieving high test coverage}\textsuperscript{(4)} and reliable testing practices are particularly challenging in the context of ML, further complicating CI implementation.
According to P66, \textit{``ML projects have a more nuanced testing requirement for CI purposes. Unit tests are not enough, we need to have data-centric tests such as data validation, verification. This would mean, checks at multiple stages of the pipeline(or workflow)''}. In addition, P57 highlights that \textit{``ML projects often are stochastic or make statistical guarantees in nature, and require different forms of testing/verification as a result''}. Furthermore, P75 shared, \textit{``we can only test the [model] performance on old data which might not be a good evaluation, whereas in non-ML projects we can just build on top of existing unit tests, in ML projects the tests can change depending on data and model''}. 

\textbf{\textit{Resource and Infrastructure Requirements}}.\textsuperscript{(43)}
This theme includes the significant resource demands and infrastructure complexities necessary for successful CI implementation in ML.
Notably, the code \textit{more computational resources required}\textsuperscript{(32)} highlights a critical barrier that teams must overcome to effectively implement CI practices in ML projects. The participants’ feedback indicates that \textit{higher resource costs}\textsuperscript{(11)} and \textit{the need for configurations and debugging of low-level libraries}\textsuperscript{10} add additional layers of complexity to the CI process. As mentioned by P118, \textit{``ML models often require significant computational resources, especially for training complex models or processing large datasets. CI pipelines need to provision and manage these resources efficiently, which may involve using specialized hardware accelerators (e.g., GPUs) or cloud-based services''}. Additionally, P96 explains that \textit{``Often, models are too large to be practically used on CPU runners, and thus require a runner to be equipped with a CUDA-enabled GPU, which most CI platforms do not provide and require users to set up \& manage their own runner''}.

\textit{\textbf{Build Duration and Stability}}.\textsuperscript{(32)}
It encompasses the challenges of managing long build durations and ensuring stability in the CI process in the ML domain.
Notably, \textit{long build durations}\textsuperscript{(28)} and \textit{increased build break proneness}\textsuperscript{(5)} emerged as the most frequently cited issues related to the build process in ML projects. These findings corroborate with our previous study that quantitatively shows that ML projects face challenges related to extended build duration \citep{bernardo2024machine}, which can hinder rapid development cycles. Participants have indicated that maintaining short build durations is a critical aspect of CI practices that is often not fully realized. 
As P40 emphasized, \textit{``Workflows are much longer since models are large and resource extensive. One could either spend a ton of money to use larger GPUs or decrease testing coverages''}. Additionally, handling the data needed to test the models also put additional complexity to the build process. P85 explained, \textit{``When huge amounts of data are involved, then it can be hard to retrieve it quickly enough, which leads to high testing times"}. Indeed, the addition of data and model handling in the CI pipeline of ML projects direct impact on their build duration. For instance, P31 shared, \textit{``When doing Unitest, I feel that it takes more time than in other projects where it is not directly related to the code (e.g. when downloading a model once on \textsc{GitHub} Workflow and taking it to test, it takes quite a long time)"}.

\textit{\textbf{Data Handling and Management}}.\textsuperscript{(30)}
This theme focuses on the complexities involved in managing data throughout the CI pipeline of ML projects, such as \textit{data dependency}\textsuperscript{(7)}, \textit{data management}\textsuperscript{(6)}, and \textit{large data volume}\textsuperscript{(5)}, showing that practitioners often struggle with ensuring data quality and accessibility. As mentioned by several participants, effective management of data is essential for successful CI implementation, highlighting the need for robust strategies to handle diverse data requirements in ML projects.
For instance, P100 explained, \textit{``ML projects often deal with large, complex datasets, requiring robust data management and handling strategies. This introduces challenges in terms of storing, accessing, and versioning datasets within the CI pipeline''}. In addition, P14 shared, \textit{``Data configuration is a complex task and needs to be solved in the pipeline too. Fixtures in this context are really complex to accomplish, but they give more value to the entire process of CI''}.

\textbf{\textit{Model Handling and Management.}}\textsuperscript{(20)} 
This theme explores the challenges of tracking and managing ML models within the CI pipeline.
Participants identified issues such as\textit{ model metrics performance tracking}\textsuperscript{(7)} and the \textit{need for ongoing model training}\textsuperscript{(6)}. These challenges underscore the importance of monitoring and maintaining models throughout their lifecycle to ensure their effectiveness, reinforcing the necessity of CI practices that can accommodate the unique demands of ML models.
As explained by P25, \textit{``There are some specific challenges that arise when implementing a continuous integration pipeline in ML projects compared to regular projects. For example, there is a need to version and store datasets and machine learning models, as well as track model performance metrics on each commit''}.
According to P128, \textit{``Sometimes keeping track of metrics for example model accuracy becomes a challenge. Especially since our output changes quite a lot during experiments''}.

\textbf{\textit{Integration and Maintenance.}}\textsuperscript{(17)} 
It highlights the difficulties in integrating CI practices with existing tools and the ongoing maintenance challenges that arise from evolving dependencies in ML projects.
The \textit{difficulty in integrating ML workloads into existing tools}\textsuperscript{(8)} and the \textit{higher maintenance effort}\textsuperscript{(4)} emphasize the challenges ML teams face when adapting CI practices to their specific environments. Participants expressed concerns regarding the \textit{instability of dependencies' APIs}\textsuperscript{(4)} and the need for \textit{additional library dependencies}\textsuperscript{(4)}, indicating that CI in ML is complicated by a constantly evolving set of tools and technologies.
P58 shared, \textit{``For builds you usually rely on many other software pieces, pytorch, hugging face etc and the datasets are sometimes large and not even publicly accessible to be able to access via a GitHub action''}. 
In addition, P49 emphasized, \textit{``Training ML models generally requires significant computational resources, such as GPUs or TPUs, which may not be readily available in standard CI environments. Integrating these resources with CI pipelines often requires additional configurations and costs''}.


\textit{\textbf{Organizational and Process Challenges}}.\textsuperscript{(8)} 
This theme presents organizational barriers that impact CI implementation in ML
projects.
The findings related to this theme reveal that factors like \textit{less mature practices}\textsuperscript{(3)} and \textit{specialized knowledge requirements}\textsuperscript{(3)} pose difficulty on the integration of CI into ML workflows. Participants noted \textit{conflicts of interest between stakeholders}\textsuperscript{(1)} and pressures from \textit{shorter deadlines}\textsuperscript{(1)}, suggesting that organizational culture and dynamics play a significant role in the adoption of CI practices in the ML domain.
For example, P20 noted, \textit{``I think CI best practices are less defined for ML projects compared to non-ML"}. Furthermore, P104 elaborated, \textit{``Domain specific knowledge required for reviewing certain features, data caching and test runtime management"}.

\textit{\textbf{Project-Specific Constraints.}}\textsuperscript{(8)}
This theme highlights challenges and differences specific to ML projects and their characteristics.
Participants emphasized the \textit{experimental nature of ML code}\textsuperscript{(3)}, the involvement of \textit{different artifacts}\textsuperscript{(2)}, and \textit{debugging hardness}\textsuperscript{(2)} are factors related to ML projects that might lead to a nuanced implementation of a CI pipeline in ML projects.
For instance, P68 noted, \textit{``Many ML project code may be temporary and experimental, it would be costly to integrate a CI pipeline"}. 
In addition, P79 remarked, \textit{``The artifacts for an ML pipeline look way different from other software projects. You’re likely testing your ML pipeline in different chunks (preprocessing, vectorizing, retraining all as separate scripts). Setting up CI to ensure each of these pieces work independently, and their outputs are correct, takes a lot more focus''}. 
Finally, a challenge associated with debugging is elucidated by P83, who stated \textit{``you have to debug some hardware errors when working on ML projects''}.

\begin{highlightbox}
\textbf{Key Findings:}
\begin{itemize}
    \item \textit{\textbf{Perceived Differences:}}     
    Most participants often or always perceive ML projects as having longer build durations (63.2\%, \nicefrac{98}{155}) and lower test coverage (61.9\%, \nicefrac{96}{155}).
    In contrast, responses regarding commit frequency and build fix times were more neutral. Most participants reported that ML projects only sometimes, rarely, or never fix broken builds more quickly (77.4\%, \nicefrac{120}{155}) or commit more frequently (74.8\%, \nicefrac{116}{155}) than non-ML projects.
    
    \item \textit{\textbf{Challenges in CI Implementation:}} Participants identified eight main themes reflecting challenges specific to ML projects:
    \textit{Testing Complexity},
    \textit{Infrastructure Requirements},
    \textit{Build Duration and Stability},
    \textit{Data Handling and Management},
    \textit{Model Handling and Management},
    \textit{Integration and Maintenance Challenges},
    \textit{Organizational and Process Challenges}, and
    \textit{Project-Specific Constraints}.
\end{itemize}
\end{highlightbox}



\subsection*{\textbf{\RQtwo}}

The results for RQ2 explore practitioners' perspectives on build duration in ML projects, focusing on their importance, expectations across project sizes, and factors contributing to longer durations.

\subsubsection*{\textbf{Practitioners' Perspectives on Build Duration: Importance and Expectations}}

\begin{figure}
	\centering
	\includegraphics[width=12cm]{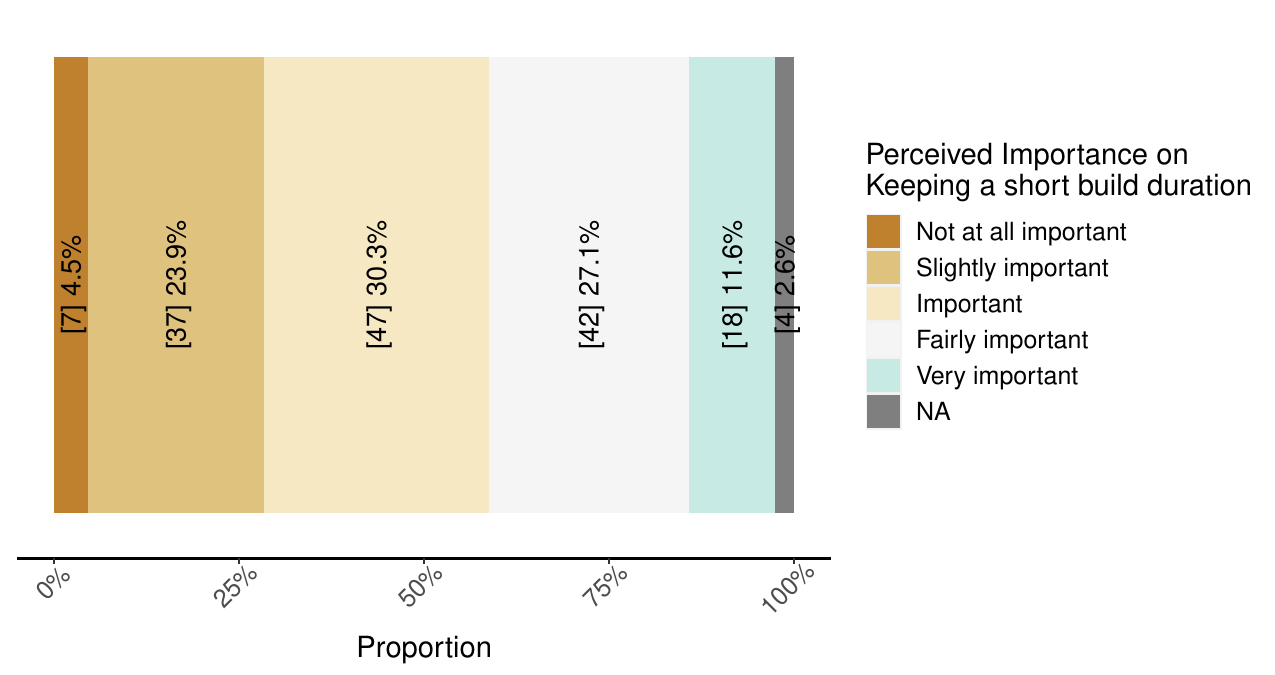}	
	\caption{Participants' perceived importance on ML projects keeping a short build duration.}	\label{fig:q3_1_Perceived_importance_on_ML_projects_keeping_a_short_build_duration}       
\end{figure}

Figure~\ref{fig:q3_1_Perceived_importance_on_ML_projects_keeping_a_short_build_duration} illustrates participants' perceptions of the importance of keeping a short build duration in ML projects. The responses are categorized into five levels of importance, ranging from \textit{``Not at all important"} to \textit{``Very important"}, with the additional \textit{``N/A''} category for the absence of a response. 
The majority of participants (69\%, \nicefrac{107}{155}) recognized the importance of keeping build durations short in ML projects. Specifically, 30.3\% (n = 47) of participants rated this factor as \textit{``Important''}, 27.1\% (n = 42) considered it \textit{``Fairly important''}, while 11.6\% (n = 18) indicated that keeping build durations short is \textit{``Very important''}.
Interestingly, 23.9\% (n = 37) of participants rated the importance of keeping a short build duration as \textit{``Slightly important"}, while 4.5\% (n = 7) considered it \textit{``Not at all important"}. These results indicate that while build duration is acknowledged as a relevant factor, some practitioners may prioritize other aspects of CI pipelines.

\begin{figure}
	\centering
	\includegraphics[width=12cm]{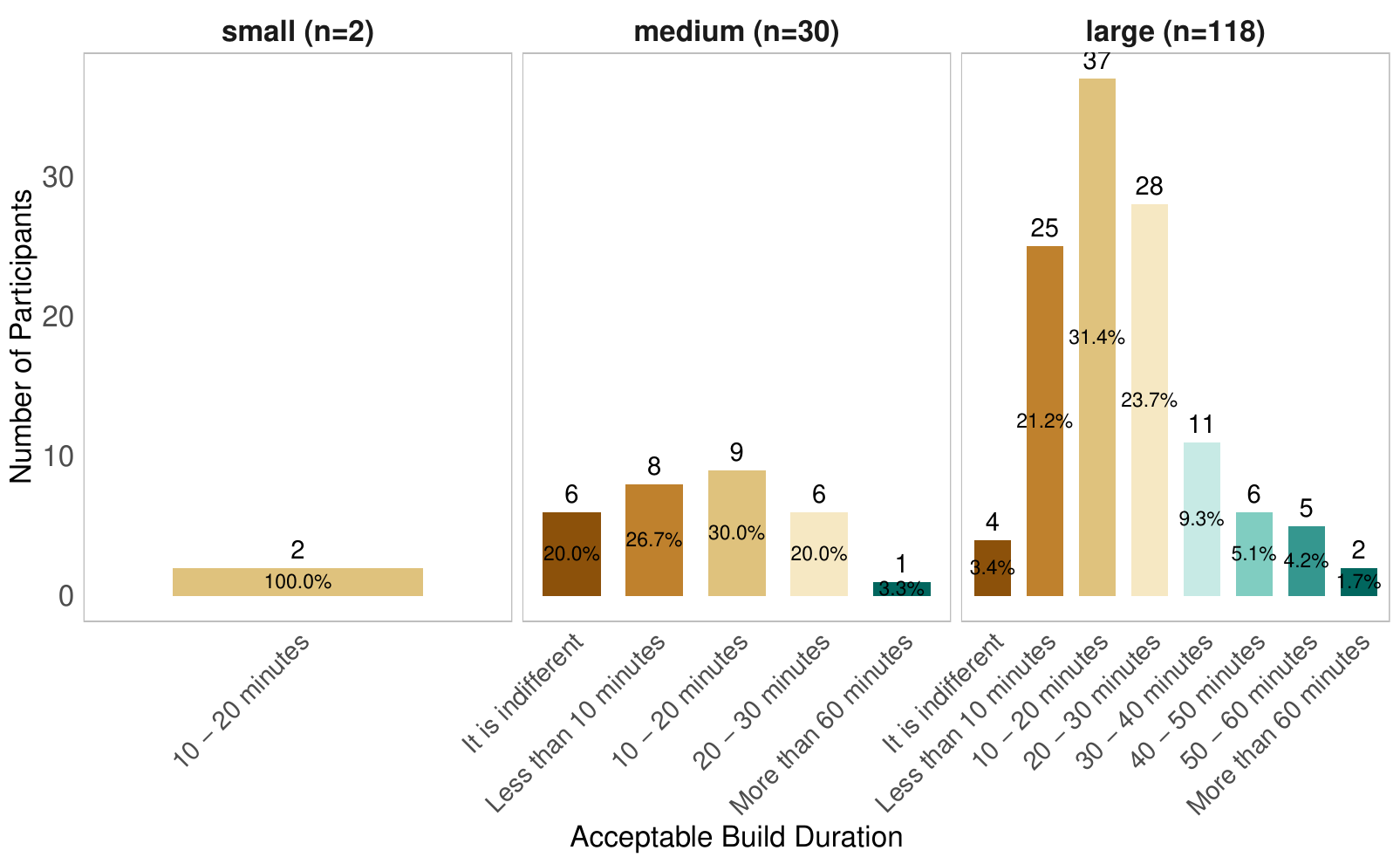}
	\caption{Participants' perceptions of acceptable build duration of ML projects, categorized by project size (small, medium, large).}	\label{fig:q4_2_Acceptable_build_duration_for_an_ML_project_of_comparable_size_to_the_studied_project}       
\end{figure}

To gain deeper insights into how ML practitioners manage build durations, we asked participants in \textsc{Question \#4.2} about their perception of an acceptable build duration for ML projects of similar size to their own. Figure~\ref{fig:q4_2_Acceptable_build_duration_for_an_ML_project_of_comparable_size_to_the_studied_project} illustrates these perceptions, categorized by project size (small, medium, and large), providing a clear overview of how acceptable build durations vary across different project scales.

For small ML projects, there is a consensus on acceptable build durations, with all participants ($100\%, n = 2$) indicating a preference for builds lasting 10–20 minutes. This aligns closely with the observed median build duration of 10.3 minutes from our prior quantitative study~\citep{bernardo2024machine}, reinforcing the emphasis on speed and efficiency in small projects. These projects tend to feature simpler workflows, fewer dependencies, and lower resource demands, which naturally support shorter build times. However, the limited sample size ($n = 2$) reduces the generalizability of this finding, requiring cautious interpretation.

In medium-sized ML projects, a wider variety of expected build durations is deemed acceptable, indicating that his group has more diverse needs. The majority of participants ($56.7\%, n = 17$) expect relatively short builds, with 26.7\% ($n = 8$) preferring durations under 10 minutes and another $30\% (n = 9)$ favoring builds lasting 10–20 minutes. These preferences align well with the observed median build duration of $12.9$ minutes from our prior work \citep{bernardo2024machine}, suggesting that practitioners generally expect builds in medium-sized projects to remain efficient while accommodating moderate increases in complexity. However, the tolerance for slightly longer durations -- such as the $20\% (n = 6)$ of participants who find builds of 20–30 minutes acceptable -- indicates that practitioners acknowledge the trade-offs between quick feedback and the growing demands of more intricate workflows. Additionally, the $20\% (n = 6)$ of participants indifferent to build duration may represent those who prioritize factors such as stability or comprehensive testing over build speed.

In case of larger ML projects, 20.3\% (n = 24) of practitioners consider build times exceeding 30 minutes acceptable, compared
to only 3.3\% (n = 1) in medium-sized projects.
While 21.2\% (n = 25) of participants prefer builds lasting less than 10 minutes, the largest group (31.4\%, n = 37) considers durations of 10–20 minutes acceptable, followed by $23.7\% (n = 28)$ who are comfortable with builds lasting 20–30 minutes. Combined, 76.3\% (n = 90) of respondents favor builds under 30 minutes, aligning closely with the observed median build duration of 21.4 minutes reported in our prior work \citep{bernardo2024machine}.
However, the acceptance of builds exceeding 30 minutes by 20.3\% (n = 24) underscores the trade-offs inherent in scaling large ML systems. This alignment between practitioners’ expectations and observed build durations reflects a nuanced understanding of the challenges associated with scaling ML projects, which often involve complex dependencies, resource-intensive computations, and extensive validation processes. 

\subsubsection*{\textbf{Factors Contributing to Longer Build Durations}}

In our previous work \citep{bernardo2024machine}, we quantitatively compared the build durations of ML and non-ML projects, finding that ML projects generally exhibit longer build times. To assess whether these findings align with practitioners' expectations, we included this topic in \textsc{Question \#3.2} of our survey.
The majority of respondents (75.5\%, 117 out of 155) agreed with our findings, indicating that they expect ML projects to have longer build durations. In contrast, 5.2\% (8 out of 155) disagreed. While many of these participants did not provide a clear rationale for their disagreement, some suggested that smaller ML projects might not experience significantly longer build times compared to non-ML projects. For example, P37 noted, \textit{``My projects are relatively small so the impact is usually not that noticeable''}.
A further 3.9\% (6 out of 155) of respondents highlighted that the build duration depends on project-specific characteristics. As P62 explained, \textit{``it [the build duration] does not only depends on LOC, but mostly on the nature of the ML technology or models to use. Is not the same to run a build using a project with some light models \& libraries like XGBoost or scikit-learn vs Deep Learning and LLMs"}. 
Finally, 17.4\% (24 out of 155) of participants did not provide an answer or a clear response to the question.

\begin{figure}
	\centering
	\includegraphics[ width=12cm]{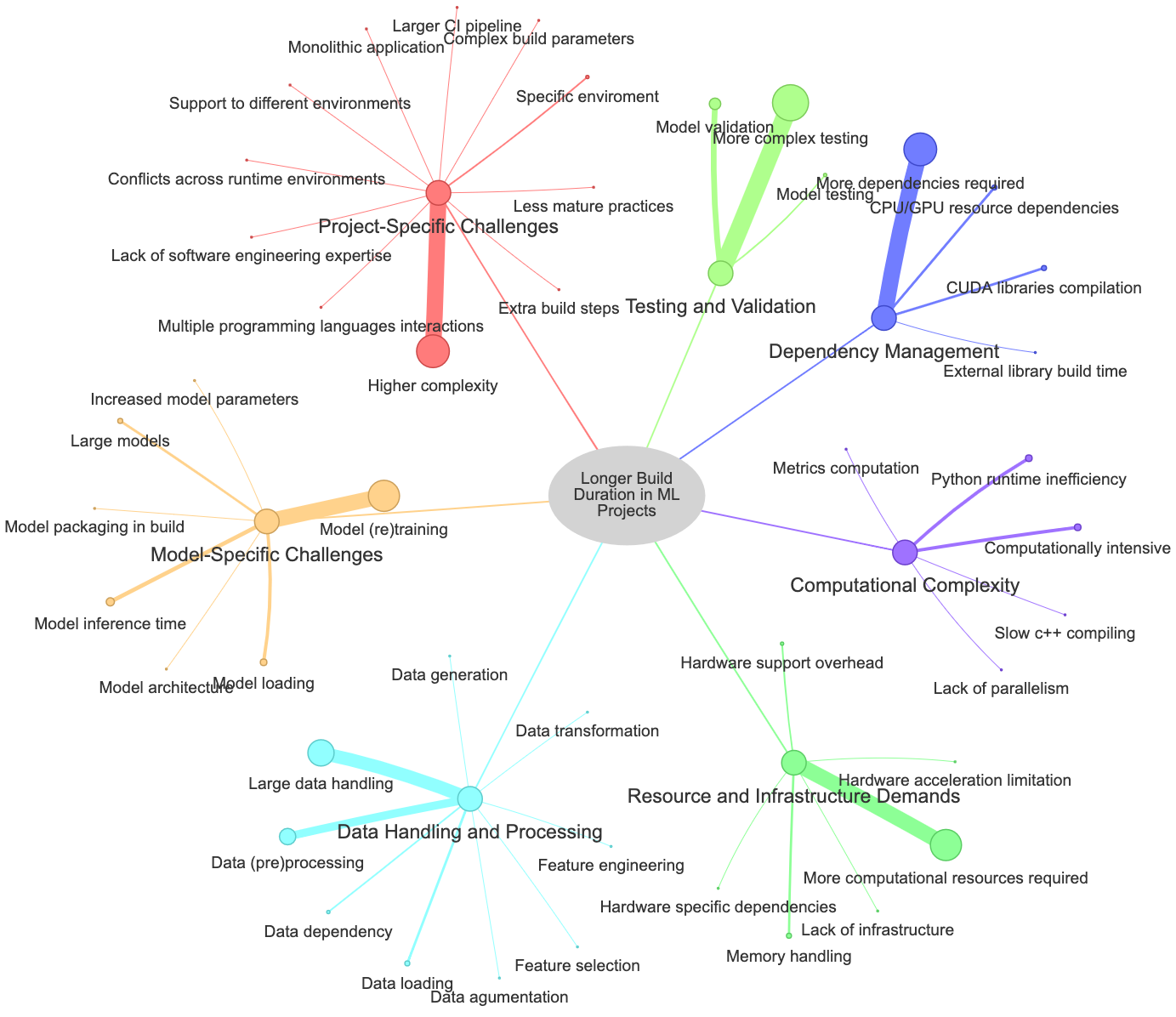}
	\caption{Perceived Reasons for Longer Build Duration in the CI pipeline of ML projects.}
	\label{fig:q3_2_reasons_for_longer_build_duration}       
\end{figure}

Participants identified 45 factors that contribute to longer build durations in ML projects, which were grouped into seven main themes. Figure \ref{fig:q3_2_reasons_for_longer_build_duration} illustrates the themes and associated codes that emerged from our thematic analysis.

The most frequently cited themes  were \textit{Project-Specific Challenges}\textsuperscript{(30)}, \textit{Model-Specific Challenges}\textsuperscript{(28)}, and \textit{Testing and Validation}\textsuperscript{(26)}. Other significant themes included \textit{Data Handling and Processing}\textsuperscript{(25)}, \textit{Resource and Infrastructure Demands}\textsuperscript{(25)}, \textit{Dependency Management}\textsuperscript{(24)}, and \textit{Computational Complexity}\textsuperscript{(11)}. These themes highlight the varied and complex factors contributing to extended build durations in ML projects.
We provide detailed explanations of the main codes linked to each theme in the following.

\textit{\textbf{Project-Specific Challenges}}.\textsuperscript{(30)} This theme captures the diverse and often unique project factors that can impact build duration in ML projects. Participants highlighted \textit{higher complexity}\textsuperscript{(20)} and \textit{specific environments}\textsuperscript{(2)} as major influences, noting that ML projects often have specialized requirements that add complexity to the CI process. As P07 explained, \textit{``ML projects require complex computations that can take long and as a result, the builds will take longer from setting up the environment for large libraries and executing complex intensive codes"}.
Issues such as \textit{complex build parameters}\textsuperscript{(1)}, \textit{conflicts across runtime environments}\textsuperscript{(1)}, and \textit{support for different environments}\textsuperscript{(1)} were also mentioned, emphasizing the challenges ML projects face in maintaining consistency across varied setups. 
P41 elaborated, \textit{``ML projects can encounter conflicts across various runtime environments as well as challenges in resource allocation. Moreover, running an ML project without thorough checks often leads to 'out of memory' errors.''}.

Additionally, some respondents pointed to \textit{monolithic application architectures}\textsuperscript{(1)}, where data sources are often tightly coupled with the model, requiring engineers to push the entire codebase as a single, monolithic application. 
Other factors cited as making builds more challenging and time-consuming in the ML domain include extra build steps\textsuperscript{(1)}, interactions between multiple programming languages\textsuperscript{(1)}, larger CI pipelines\textsuperscript{(1)}, as well as a lack of software engineering expertise\textsuperscript{(1)} and less mature development practices\textsuperscript{(1)}.

\textit{\textbf{Model-Specific Challenges}}.\textsuperscript{(28)} This theme reflects the unique demands associated with ML models, which often require additional processing steps that lengthen build times. For example, \textit{model (re)training}\textsuperscript{(19)} was frequently mentioned, as many ML pipelines necessitate frequent retraining to ensure model accuracy. Other codes within this theme include \textit{model inference time}\textsuperscript{(5)}, \textit{model loading}\textsuperscript{(4)}, and \textit{handling large models}\textsuperscript{(3)}, each contributing to increased build durations. As highlighted by P134, \textit{``While a short build time is generally desirable, the need for more data processing as well as involving basic training and inference will likely increase the build time''}.
Other participants noted the difficulty of integrating model-related processes into the CI pipeline, with factors like \textit{model architecture}\textsuperscript{(1)}, \textit{model packaging in builds}\textsuperscript{(1)}, and \textit{increased model parameters}\textsuperscript{(1)} posing additional hurdles that are not typically encountered in non-ML projects.
P31 explained, \textit{``A large ML project would include multiple functions (or multiple models) and the number of parameters in the model would increase accordingly''}.

\textit{\textbf{Testing and Validation}}.\textsuperscript{(26)} This theme focuses on the extensive testing requirements in ML projects, which add significant time to the build process. \textit{More complex testing}\textsuperscript{(22)} was the most frequently cited factor, as ML projects often require not only traditional software testing but also additional validation steps specific to model performance and data quality. 
\textit{Model validation}\textsuperscript{(7)} and \textit{model testing}\textsuperscript{(2)} were also highlighted, with respondents noting that these processes are essential to ensure reliable model behavior and accurate results but can considerably extend build times. As P57 explained, \textit{``ML projects often have multiple steps that need to happen over (even in test settings) non-trivial data sizes to validate correctness. E.g., a full integration test for a project I'm involved needs to extract some synthetic raw data from disk, re-format it, pre-process it, then ``train'' a model on a synthetic task over that data to convergence, then validate that model. Each of these steps use real tools designed for production use (so that the test covers appropriate, real-world usage) and the end to end pipe takes a long time''}.

\textit{\textbf{Data Handling and Processing}}.\textsuperscript{(25)} This theme represents the complexities involved in managing large datasets, a core aspect of many ML projects that significantly impacts build duration. Participants pointed to \textit{large data handling}\textsuperscript{(16)} and \textit{data preprocessing}\textsuperscript{(10)} as two of the most time-consuming processes. \textit{Data loading}\textsuperscript{(3)} and \textit{data dependency management}\textsuperscript{(2)} were also noted as factors adding complexity, with one respondent explaining that handling large volumes of data slows down the CI pipeline, especially when specific dependencies or transformations are required. Other aspects such as \textit{data augmentation}\textsuperscript{(1)}, \textit{data generation}\textsuperscript{(1)}, \textit{data transformation}\textsuperscript{(1)}, \textit{feature engineering}\textsuperscript{(1)}, and \textit{feature selection}\textsuperscript{(1)} highlight the extensive data manipulation often required in ML builds, where preparing data for model training and validation can be as demanding as model-related processes. As P100 emphasized,
\textit{``ML projects inherently introduce additional computational burdens that contribute to longer build times. This is especially evident in medium and large ML projects, where the data processing and model training tasks become more significant. ML projects typically involve extensive data handling, including loading, cleaning, transforming, and preprocessing. These steps can be computationally intensive, especially when dealing with large datasets. This contributes to longer build durations''}.

\textit{\textbf{Resource and Infrastructure Demands}}.\textsuperscript{(25)} This theme addresses the substantial resource requirements that ML projects often impose, which can prolong build times when adequate infrastructure is not readily available. The need for \textit{more computational resources}\textsuperscript{(19)} was a recurrent point, with participants emphasizing the requirement for powerful hardware like GPUs and TPUs. P10 noted, \textit{``from my experience ML projects tend to have much longer build times since they utilize far more CPU/GPU resources than non-ML projects''}. In addition, P25 explained, \textit{``I've noticed that projects with machine learning tend to have build times that are several times longer than projects without ML. This is especially noticeable as the project size and complexity increase. I think this is due to the fact that ML projects require more computational resources and time for tasks specific to ML, such as data preprocessing, model training, and evaluation. These tasks can be very time-consuming, especially as the size of the project and dataset grows''}. 

Limited access to specialized hardware can delay builds, as noted by respondents who mentioned \textit{memory handling}\textsuperscript{(3)}, \textit{hardware support overhead}\textsuperscript{(2)}, \textit{hardware acceleration limitations}\textsuperscript{(1)}, \textit{hardware specific dependencies}\textsuperscript{(1)}, and the \textit{lack of infrastructure}\textsuperscript{(1)} as constraints. Practitioners expressed difficulties in securing the necessary resources to maintain efficient CI pipelines in ML projects. As explained by P49, \textit{``ML projects can encounter conflicts across various runtime environments as well as challenges in resource allocation. Moreover, running an ML project without thorough checks often leads to 'out of memory' errors''}.


\textit{\textbf{Dependency Management}}.\textsuperscript{(24)} This theme highlights the challenges posed by the large number of dependencies in ML projects, as well as the extensive data and parameters required by specialized ML libraries during installation, which can introduce significant build overhead. Since CI pipelines typically start from a fresh environment, these dependencies must be reinstalled in every run, further slowing down the process.
\textit{More dependencies required}\textsuperscript{(20)} was frequently cited, alongside issues related to \textit{CPU/GPU resource dependencies}\textsuperscript{(3)} and the \textit{compilation of CUDA libraries}\textsuperscript{(3)}, both of which are common in ML projects that rely on hardware acceleration. Participants also mentioned the time-consuming process of managing \textit{external library build times}\textsuperscript{(1)}, with one respondent pointing out that ML projects often depend on complex libraries that require specific configurations, making builds slower and more resource-intensive. As P94 observed, \textit{``ML projects tends to have a lot of dependencies, a large portion of them being hardware specific''}. Similarly, P63 explained, \textit{``I think it's [longer build duration] because of the various dependencies [ML projects require], for example if you want to any DL framework you need to install cuda which takes time, if you have a repo of various ML models ranging from text to vision they will have their long list of dependencies, so I think it's because of the various dependencies ML projects take much longer to build''}.

\textit{\textbf{Computational Complexity}}.\textsuperscript{(11)} It highlights the resource-intensive nature of ML projects, which often require substantial computational power and face efficiency challenges that extend build times. Participants frequently cited \textit{computationally intensive processes}\textsuperscript{(4)} and \textit{Python runtime inefficiencies}\textsuperscript{(4)} as factors slowing down builds. 
As P96 explained, \textit{``The computational requirements of machine learning pipelines intuitively takes significantly more time than the relatively quick unit \& integration tests of traditional software''}.
Limited parallel processing was also mentioned, with some respondents pointing to a \textit{lack of parallelism}\textsuperscript{(1)} and the impact of \textit{metrics computation}\textsuperscript{(1)} as significant contributors to extended durations. Additionally, \textit{slow C++ compilation}\textsuperscript{(1)} was noted, particularly in projects where C++ is used for performance-critical components. 
As P144 observed, \textit{``Most of the ML stack is written in Python and C++. The C++ compiler is pretty slow + rebuilding the stubs''}.
Together, these issues reflect the intensive computational demands that make rapid builds more challenging in ML projects. 

\begin{highlightbox}
\textbf{Key Findings:}
\begin{itemize}
    \item \textit{\textbf{Importance of Build Duration:}} 69\% of participants recognize short build durations as important, though this perception trend changes with project size and complexity.
    \item \textit{\textbf{Project Size Variability:}} Small ML projects tend to favor build times of 10–20 minutes.
    In medium-sized projects, most practitioners (56.7\%) also prefer short build times (10-20 minutes), while 20\% find durations up to 30 minutes acceptable.
    In contrast, large-sized projects are more tolerant of longer builds, with 20.3\% of practitioners considering build times exceeding 30 minutes acceptable, compared to only 3.3\% in medium-sized projects.
    \item \textit{\textbf{Expectation of Longer Builds:}} 75\% of participants expect longer build durations in ML projects compared to non-ML projects, driven by their complexity.
    \item \textit{\textbf{Reasons for Longer Builds:}} Participants highlighted seven key themes contributing to longer build durations in ML projects. The most frequently mentioned reasons include \textit{higher project complexity}, \textit{model training} during builds, \textit{increased computational resource demands}, \textit{extensive data handling}, and \textit{the need for managing numerous dependencies}.
\end{itemize}
\end{highlightbox}

\subsection*{\textbf{\RQthree}}

The results for RQ3 examine practitioners' perceptions of the challenges in testing ML projects and their views on acceptable test coverage rates for these projects.

\subsection*{\textbf{Challenges in Testing ML Projects}}

In \textsc{Question \#3.6} of our survey, participants were asked about the perceived challenges ML projects face when testing their source code, which we subsequently analyzed through a thematic analysis. 

\begin{figure}
	\centering
	\includegraphics[ width=12cm]{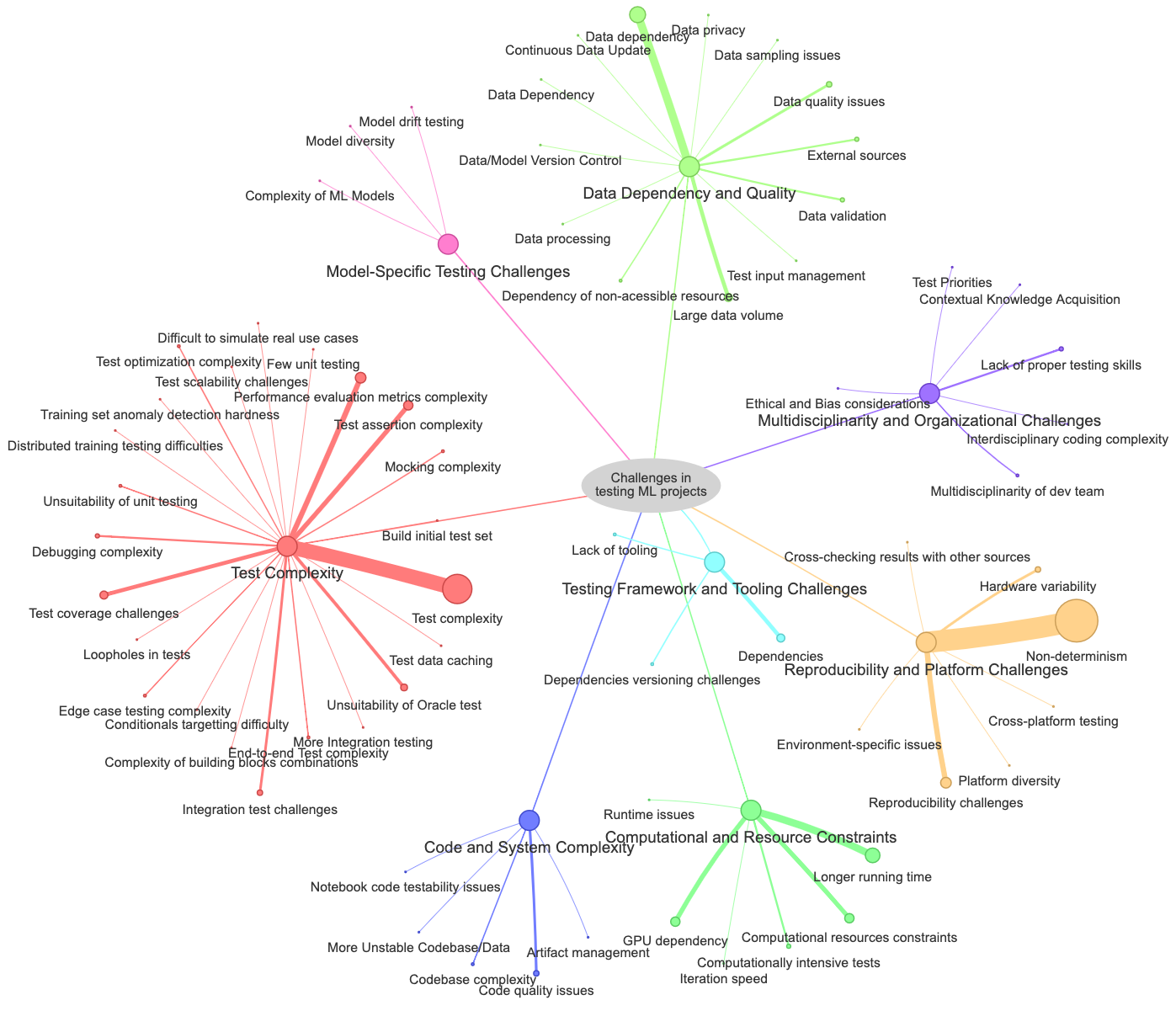}
	\caption{Perceived Challenges in Testing ML projects.}
	\label{fig:q3_6_challenges_in_testing_ml_projects}       
\end{figure}

Our participants identified 66 factors that can pose challenges to testing source code in ML projects.
These factors were categorized into eight key themes, as illustrated in
Figure~\ref{fig:q3_6_challenges_in_testing_ml_projects}, which presents the themes and their associated codes derived from our thematic analysis.

The most frequently cited themes were \textit{Test Complexity}\textsuperscript{(56)}, \textit{Reproducibility and Platform Challenges}\textsuperscript{(41)}, \textit{Data Dependency and Qualit}y\textsuperscript{(31)}, and \textit{Computational and resource Constraints}\textsuperscript{(26)}. Other themes included
\textit{Testing Framework and Tooling Challenges}\textsuperscript{(9)},
\textit{Code and System Complexity}\textsuperscript{(8)},
\textit{Multidisciplinarity and Organizational Challenges}\textsuperscript{(7)} and
\textit{Model-Specific Testing Challenges}\textsuperscript{(3)}. These themes highlight the varied and complex factors that challenge the testing process of ML projects. 
We provide detailed explanations of the main codes linked to each theme in the following.

\textbf{\textit{Test Complexity.}}\textsuperscript{(56)}
This theme emerged as a recurrent challenge, with the highest number of references. Practitioners frequently cited the overall \textit{test complexity}\textsuperscript{(22)}, emphasizing the intricate nature of crafting, executing, and maintaining tests for ML systems. Unlike traditional software, ML projects often involve probabilistic outcomes, making it difficult to define deterministic assertions, as reflected by challenges such as \textit{test assertion complexity}\textsuperscript{(7)} and \textit{unsuitability of Oracle tests}\textsuperscript{(5)}.
For instance, P48 noted, ``\textit{ML models often produce different results on different runs due to inherent randomness in training processes. Traditional unit tests focus on code correctness, but ML models require validation of performance metrics (e.g., accuracy, precision, recall) on test datasets, which is more complex and less binary than typical pass/fail unit tests''}.

Furthermore, issues like \textit{test coverage challenges}\textsuperscript{(6)} and \textit{integration test challenges}\textsuperscript{(4)} indicate that achieving comprehensive coverage across diverse system components remains a significant obstacle.
P101 noted, \textit{``ML projects need higher test case coverage, which is difficult, because there can be a lot of possible paths"}. Additionally, P114 shared, \textit{``Writting in test shapes/scenarios that offer complete coverage is difficult. Traditionally we can measure coverage through code paths posted, but this breaks down when different data strategies might hit the same API's (and cause bugs in some cases)''}.
Performance evaluation in ML projects also presents a unique challenge, with \textit{performance evaluation metrics complexity}\textsuperscript{(8)} highlighting the difficulty of relying on metrics such as precision and recall rather than traditional binary outcomes. P138 explained, \textit{``It's not easy to implement a test which decides e.g. whether model achieves a certain value of performance metric on a popular dataset''}.
Additional difficulties include debugging complex systems, handling edge cases, and ensuring scalability, further complicating the testing process.

\textbf{\textit{Reproducibility and Platform Challenges.}}\textsuperscript{(41)}
This theme encompasses the difficulties in ensuring consistent and reproducible test results in ML projects. A significant issue is \textit{non-determinism}\textsuperscript{(32)}, which stems from factors such as random weight initialization and stochastic training processes. This inherent unpredictability complicates the ability to achieve consistent test outcomes. For instance, P118 observed, \textit{``ML models often produce non-deterministic outputs due to randomness in algorithms (e.g., random initialization of weights in neural networks) or variations in training data. This makes it challenging to validate outputs consistently across different test runs''}.
Another key aspect is the broader \textit{reproducibility challenges}\textsuperscript{(8)}, which are further intensified by issues such as \textit{hardware variability}\textsuperscript{(4)} and \textit{platform diversity}\textsuperscript{(1)}. Variations in execution environments can lead to inconsistencies in test results across systems. Additional platform-related issues, such as \textit{cross-checking results with other sources}\textsuperscript{(1)} and \textit{cross-platform testing}\textsuperscript{(1)}, further underscore the unique challenges faced by ML projects in comparison to non-ML systems.
As P89 explained, \textit{``ML models are not reproducible so designing tests that work across multiple hardware platforms is hard''}.

\textbf{\textit{Data Dependency and Quality.}}\textsuperscript{(31)}
This theme illustrates the critical role of data in ML testing. \textit{Data dependency}\textsuperscript{(12)} was identified as a major challenge, reflecting the reliance on high-quality and representative datasets to ensure meaningful test results. 
As P119 emphasized, \textit{``ML algorithms often need to generate data to validate algorithms. Thinking about the data and assertions for validating algorithms is more involved compared to non-ML projects''}.
Challenges related to \textit{large data volumes}\textsuperscript{(6)} further complicate testing, as processing and managing such datasets demand significant resources. P98 explained, \textit{``A lot of data is required to verify even simple models and pipelines''}.
Issues of \textit{data quality}\textsuperscript{(4)}, \textit{data validation}\textsuperscript{(3)}, and dependence on \textit{external sources}\textsuperscript{(3)} highlight the challenges in ensuring reliable test inputs. 
P10 noted, \textit{``High-quality data is crucial for ML models. Unclean or noisy data can lead to inaccurate predictions''}.
Furthermore, practitioners noted specific issues such as \textit{dependency on non-accessible resources}\textsuperscript{(2)}, \textit{continuous data updates}\textsuperscript{(1)}, and \textit{data privacy}\textsuperscript{(1)}, which illustrate the complexities of handling data pipelines. These challenges are unique to ML projects, where data plays a more central role in testing compared to traditional software systems.

\textbf{\textit{Computational and Resource Constraints.}}\textsuperscript{(26)}
This theme reflects the resource-intensive nature of ML testing. Practitioners highlighted the challenge of \textit{longer running times}\textsuperscript{(11)} due to the computational demands of training and testing ML models. Coupled with \textit{GPU dependency}\textsuperscript{(7)} and \textit{computational resources constraints}\textsuperscript{(7)}, these challenges significantly limit testing efficiency and scalability.
P21 explained, \textit{``Some code can ONLY run on GPUs so is very hard to test in a CI context. Some code requires lots of data to be tested on an integration level"}.
The need for substantial resources for tasks such as \textit{computationally intensive tests}\textsuperscript{(3)} and the resulting delays in \textit{iteration speed}\textsuperscript{(1)} create additional barriers to achieving comprehensive test coverage. 
As P49 explained, \textit{``ML systems often need to scale to handle large volumes of data. Testing how these systems scale, not just in terms of data size but also with respect to the computational and memory resources, is critical and challenging''}.
These constraints are not commonly seen in non-ML projects, making this a distinctly ML-specific challenge.

\textbf{\textit{Testing Framework and Tooling Challenges.}}\textsuperscript{(9)}
Practitioners frequently highlighted significant gaps in the existing tooling ecosystem for ML testing. \textit{Dependencies}\textsuperscript{(6)} and \textit{dependencies versioning challenges}\textsuperscript{(2)} were among the most commonly cited issues, reflecting the inherent complexity of managing extensive dependency trees in ML projects. As P55 explained, ML projects often require \textit{``more data input and more dependencies''}, which adds to the intricacy of the testing process.
Similarly, P140 noted that \textit{``integration and system tests are often hard as they require a number of pipeline steps and dependencies''}, emphasizing the challenges posed by the interdependencies of various components. Furthermore, the \textit{lack of tooling}\textsuperscript{(2)} specifically designed for ML testing underscores a critical gap in the current landscape, further complicating efforts to achieve adequate test coverage. As P51 observed, \textit{``lack of tooling, artifact management, data processing and dependencies can all be challenges''} in effectively testing ML projects.

\textbf{\textit{Code and System Complexity.}}\textsuperscript{(8)}
Challenges related to code and system complexity were notably prominent in ML projects. These systems often involve unstable and intricate codebases, as reflected by issues such as \textit{code quality issues}\textsuperscript{(4)}, \textit{codebase complexity}\textsuperscript{(2)}, and the presence of \textit{more unstable codebase/data}\textsuperscript{(1)}.
As explained by P100, \textit{``ML source code includes large amounts of numerical methods related to optimization and linear algebra routines. Obviously, whether this is really a difficulty depends on a person's background, but overall, there are few people with the necessary background to be able to skillfully navigate and contribute to a large ML codebase''}.
Unique challenges, such as testing \textit{notebook code}\textsuperscript{(1)}, further distinguish ML workflows from those of non-ML projects. As P54 observed, \textit{``Code quality [in ML projects] is generally worse, for many reasons''}. Similarly, P82 noted, \textit{``Lots of notebook code, lower quality, not very testable''}.
These complexities make testing ML systems fundamentally different from non-ML projects.

\textbf{\textit{Multidisciplinarity and Organizational Challenges.}}\textsuperscript{(7)}
This theme highlights the need for interdisciplinary expertise in ML testing. Challenges such as the \textit{lack of proper testing skills}\textsuperscript{(3)} and the \textit{multidisciplinarity of development teams}\textsuperscript{(2)} emphasize the importance of collaboration between software engineers, data scientists, and domain experts.
As P78 explained, the main challenge in testing ML projects is \textit{``asking an ML developer whose expertise is writing ML code to write tests (which is not his strength)''}. Similarly, P36 observed, \textit{``AI scientists not having good experience testing, software engineers not having good enough AI knowledge''}.
Additional considerations, such as \textit{ethical and bias considerations}\textsuperscript{(1)} and \textit{contextual knowledge acquisition}\textsuperscript{(1)}, further underscore the organizational and knowledge-based complexities unique to ML testing. 

\textbf{\textit{Model-Specific Testing Challenges.}}\textsuperscript{(3)}
This theme highlights the unique aspects of testing ML models. Challenges such as the \textit{complexity of ML models}\textsuperscript{(1)}, \textit{model diversity}\textsuperscript{(1)}, and \textit{model drift testing}\textsuperscript{(1)} illustrate the dynamic and evolving nature of ML systems. 
For instance, when discussing model complexities, P70 explained, \textit{``Our code needs expensive GPUs to run, so the tests would need GPUs to test stuff. Also... the data is enormous \& the models dynamics are complicated. So all of the huge math and huge data aspects, as well as the GPU aspects, are special''}. Similarly, P63 noted that \textit{``model flakiness and model variety''} present significant challenges in testing ML projects. Additionally, P118 highlighted the issue of concept drift, stating, \textit{``ML models can experience concept drift, where the relationships in the data change over time, affecting model performance. Continuous monitoring and testing for model drift require specialized techniques and tools beyond traditional unit or integration testing''}.
These challenges require specialized approaches that go beyond traditional testing methodologies.

\subsection*{\textbf{Practitioner' Perception of Acceptable Test Coverage in ML Projects}}

\begin{figure}
	\centering
	\includegraphics[width=12cm]{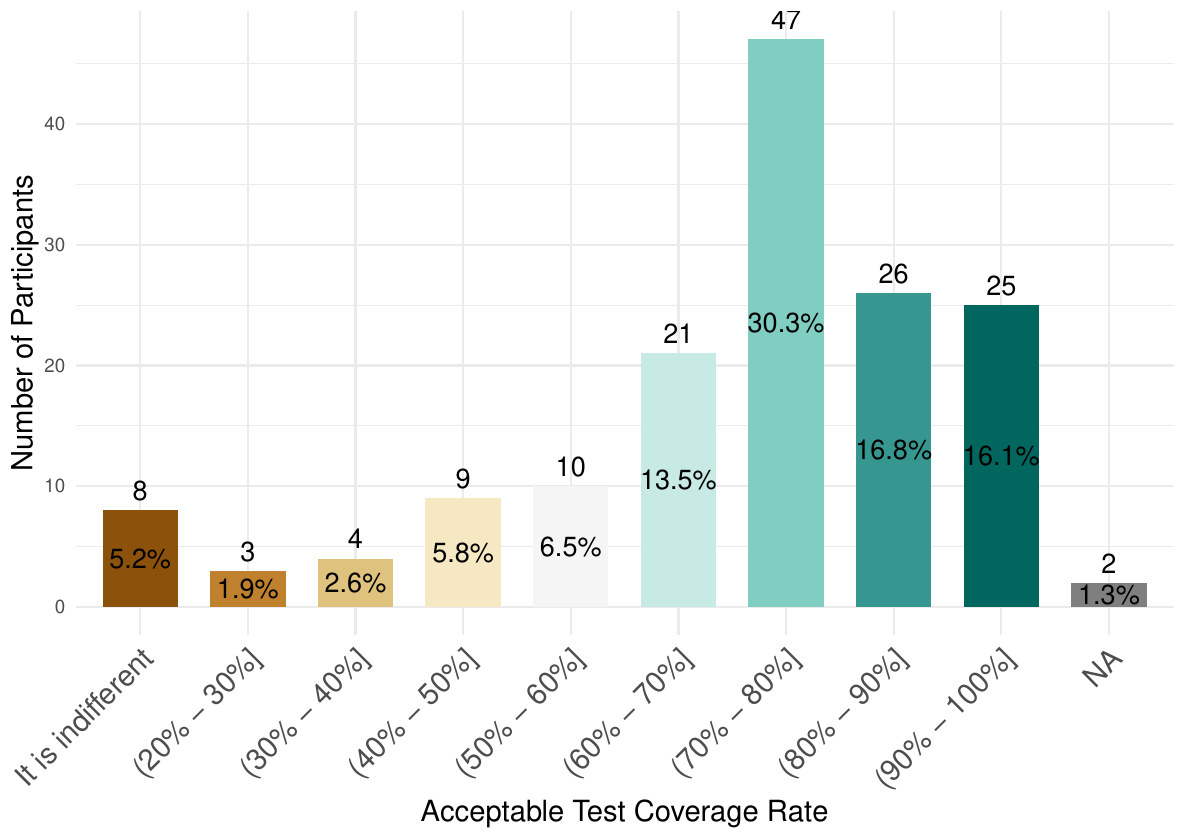}	
	\caption{Participants' perceived acceptable test coverage for ML projects.}	\label{fig:q3_4_Acceptable_test_coverage_rate_for_an_ML_project}       
\end{figure}

In \textsc{Question \#3.4}, participants were asked about their perceptions of acceptable test coverage rates for ML projects. The survey results offer valuable insights into practitioners' expectations, highlighting how these expectations align with the challenges identified in the thematic analysis. Figure \ref{fig:q3_4_Acceptable_test_coverage_rate_for_an_ML_project} visualizes the participants' responses, offering a clear perspective on acceptable test coverage rates in ML projects. Additionally, in \textsc{Question \#3.5}, participants were asked whether the result found in our prior work \citep{bernardo2024machine}, which observed that medium-sized ML projects exhibit lower test coverage rates compared to non-ML projects aligns with their expectations and to explain why they believe this discrepancy exists. 

The survey reveals that practitioners generally hold high expectations for test coverage in ML projects, with the majority indicating that acceptable coverage rates fall within the 70–80\% range (30.3\%), the 80–90\% range (16.8\%), or even the 90–100\% range (16.1\%). 
At the same time, a notable proportion of respondents viewed moderate test coverage rates as acceptable, with 13.5\% selecting 60–70\% and 6.5\% choosing 50–60\%. This reflects a pragmatic recognition of the constraints associated with ML testing. For instance, P7 explained, \textit{``Complete coverage would require resources and longer build times, so it has to be traded off''}. Similarly, P10 noted, \textit{``It's generally harder, more time-consuming, and costly to test ML projects, especially due to the intensive resource requirements and unstable results that come with ML projects compared to non-ML projects''}. These responses suggest that while practitioners aspire to high coverage, they are aware of the limitations imposed by ML-specific challenges.

Lower test coverage rates (i.e., below 50\%) were deemed acceptable by only a small fraction of participants. Just 1.9\% and 2.6\% of respondents viewed 20–30\% and 30–40\% coverage, respectively, as sufficient, while 5.8\% found 40–50\% coverage acceptable. 
The qualitative responses provide insights into why lower test coverage rates were acceptable. A key factor frequently mentioned was the inherent unpredictability of ML systems. As P136 noted, \textit{``It's not easy to write unit tests for ML projects since usually there's no deterministic expected results''}. 

Organizational and cultural factors play a significant role in shaping coverage expectations. Many participants noted that ML practitioners often prioritize experimentation over rigorous testing practices. As P113 explained, \textit{``ML engineers who write ML code are more focused on the science and algorithms, not very focused on software engineering practices such as coverage and fast-optimized builds''}. Another added, \textit{``ML projects often care more about the ability instead of bug-free code'' (P109)}. These perspectives highlight the experimental nature of ML projects, where the focus is often on achieving functional outcomes rather than building a comprehensive testing infrastructure.

Indifference to test coverage rates was reported by 5.2\% of participants. Several factors emerged from their responses to explain this perspective, including shorter project deadlines, a lower priority placed on testing, and the inherently experimental nature of many ML projects. These factors contribute to the challenges in maintaining high test coverage rates in ML workflows.
For example, P129 explained
\textit{``One of the reasons [for indifference to test coverage in ML projects] - as this topic is hot, there are more incentives to go faster and not betters''}. Similarly, P109 emphasized the focus on functionality over quality, noting, \textit{``ML projects often care more about the ability instead of bug-free code''}. P26 remarked on the difficulty of testing in ML environments, stating, \textit{``Testing in ML is hard, and people will opt not to do it if they can''}.

Additionally, some participants questioned the relevance of test coverage as a metric in the context of ML projects. P30 commented, \textit{``I am not a fan of this metric for an ML project''}, while P36 highlighted the research-driven nature of many ML workflows, observing, \textit{``ML is a lot of RnD, so lots of code is scripting that isn’t as important to test. Also, AI scientists are less likely to care about good test coverage''}.
Finally, the fast-paced and evolving nature of the ML field was identified as another factor contributing to lower test coverage rates. P59 explained, \textit{``The industry is moving fast, so communities need to deploy new technologies and stay at the state-of-the-art model. I think this is good in some ways, but it leads to projects with lower test coverage''}. These insights illustrate the practical constraints and priorities that influence practitioners' attitudes toward test coverage in ML projects.

The findings from this study can be compared to our previous research \citep{bernardo2024machine}, which revealed that medium-sized ML projects tend to have lower test coverage rates (83\%) compared to non-ML projects (94\%). While most practitioners expect high test coverage, the thematic analysis highlights the unique challenges that medium-sized ML projects face in achieving these rates. Participants frequently attributed this discrepancy to factors such as resource constraints, stochastic behaviors, the complexity of ML pipelines, and the lack of well-defined testing practices tailored to ML systems.

\begin{highlightbox}
\textbf{Key Findings:}
\begin{itemize}
    \item \textbf{\textit{Unique Challenges of ML Testing:}} Participants frequently cited test complexity, the non-deterministic nature of ML systems, data dependency, and computational resource constraints as key factors that challenge testing in ML projects.


    \item \textbf{\textit{High but Pragmatic Coverage Expectations:}} While most practitioners (63\%) favored coverage rates of 70–100\%, a significant number (20\%) accepted moderate coverage (50–70\%), citing ML-specific constraints such as resource demands and the stochastic nature of ML algorithms.



    \item \textbf{\textit{Lower Coverage in Medium-Sized ML Projects:}} 
    Participants' feedback aligned with prior quantitative findings from our earlier study \citep{bernardo2024machine}, which showed that medium-sized ML projects have lower test coverage (83\%) compared to non-ML projects (94\%). This discrepancy was attributed to factors such as shorter deadlines, the experimental nature of ML projects, and limited emphasis on rigorous testing practices.
\end{itemize}
\end{highlightbox}
\section{Discussion}
\label{sec:discussion}

In this section, we discuss the findings presented in Section \ref{sec:results}, structuring our discussion into two key topics: rethinking CI practices for ML projects (Section \ref{sec:disc_rethinking_ci_practices}) and managing build duration and test coverage in ML workflows (Section \ref{sec:handling_build_duration_and_test_cov}).


The responses from participants underscores the significance of this research in addressing the unique challenges of CI in ML projects. For example, P25 noted, \textit{``Your research is quite interesting and I'm curious to see what conclusions you ultimately reach. For ML projects today, there is often a lack of optimization and testing, so insights in this area would be very valuable.''} Similarly, P63 expressed enthusiasm for actionable findings, stating that \textit{``I really want to know about the findings!!! And would love to participate more in the future surveys''}. These perspectives highlight a genuine need for practical solutions that can help ML practitioners navigate the complexities of CI adoption. 

Beyond addressing technical obstacles, our research also resonated with participants in terms of depth and relevance. As P144 remarked, \textit{``I get a lot of these surveys, and this is one of the best ones I've taken. Really liked the analysis you did, and concrete follow-up questions. Good stuff!''} 
Such feedback reinforces the importance of translating these findings into actionable recommendations, ensuring that research-driven insights can bridge the gap between academia and real-world CI practices in ML workflows.

\subsection{Rethinking CI practices for ML projects}
\label{sec:disc_rethinking_ci_practices}

Our findings from $RQ1$ reveal key differences in CI adoption between ML and non-ML projects, primarily driven by inherent non-determinism, resource and infrastructure requirements, data dependency, and the need to accommodate model performance tracking in the CI pipeline. These challenges align with previous research. For instance, Nascimento et al. \citep{nascimento2020software} highlight the difficulty of defining correctness criteria for ML outputs due to their non-deterministic behavior. Breck et al. \citep{breck2017ml} emphasize that ML system behavior is strongly tied to data and models, which cannot always be predetermined. Additionally, Giray \citep{giray2021software} highlights the lack of mature CI testing techniques and tools for ML systems.

Insights from our survey participants reinforce these challenges, underscoring the need to rethink CI practices and develop strategies specifically tailored to ML workflows.

\subsection*{\textbf{\textit{Implication 1: ML practitioners should incorporate CI practices tailored to the ML domain (e.g., track and log model performance metrics on each commit).}}}


Although traditional software engineering has well-established CI guidelines~\citep{duvall2007continuous, fowler-ci-2006}, the ML domain still lacks standardized best practices. Recent studies have begun addressing this gap by proposing CI/CD strategies tailored for ML workflows.
Sato et al. \citep{Fowler2019CD4ML} introduce Continuous Delivery for Machine Learning (\textsc{CD4ML}), which focuses on practices to foster reproducible model training, validation data management, and model quality assurance. Similarly, Bangai et al. \citep{bagai2024implementing} explore CI/CD strategies for ML in cloud environments, highlighting key practices such as automated testing, validation, version control, and containerization.
However, despite these initial efforts, many ML practitioners still perceive a lack of well-defined CI practices in the ML domain, highlighting the need for further research and standardization. As P20 in $RQ1$ noted, \textit{``CI best practices are less defined for ML projects compared to non-ML''}. This gap underscores the importance of refining and formalizing CI principles to better support ML development workflows.

To address this gap, we draw on insights from our survey participants to define and discuss five CI practices specifically tailored to the ML domain.
Table~\ref{tab:ci_ml_practices} presents the CI practices proposed by Duvall et al. \citep{duvall2007continuous} and Fowler \citep{fowler-ci-2006}, alongside our proposed ML-specific CI practices. 
While standard CI principles such as frequent commits and quick fixes of broken builds remain applicable, ML projects introduce unique challenges that require specialized adaptations.
In the following sections, we describe each ML-specific CI practice we propose, providing explanations and supporting participant insights.

\begin{table}[]
\centering
\caption{Comparison of CI practices from Duvall, Fowler, and CI-specific practices for ML projects.}
\label{tab:ci_ml_practices}
\renewcommand{\arraystretch}{1.2} 
\begin{tabular}{>{\arraybackslash}p{2.0cm}>{\arraybackslash}p{2.5cm}>{\arraybackslash}p{2.8cm}>{\arraybackslash}p{2.8cm}}
\toprule
\textbf{Practice Category} & 
\textbf{Duvall et al. practices (\cite{duvall2007continuous})} & 
\textbf{Fowler CI Practices (\cite{fowler-ci-2006})} & 
\textbf{CI Practices for ML} \bigstrut\\
\midrule
\multirow[t]{2}{*}{\parbox[t]{2cm}{\textbf{Code Integration and Repository Management}}} & 
Commit code frequently & Everyone commits to the mainline every day & 
- \bigstrut\\
\cmidrule{2-4}

& Don’t commit broken code & Maintain a single source repository & - \bigstrut\\
\midrule
\multirow[t]{8}{*}{\parbox[t]{2cm}{\textbf{Build Management}}} & 
Run private builds & Every commit should build the mainline on an integration machine & - \bigstrut\\
\cmidrule{2-4}

& Fix broken builds immediately & Fix broken builds immediately & - \bigstrut\\
\cmidrule{2-4}

& - & Automate the build & - \bigstrut\\
\cmidrule{2-4}

& - & Keep the build fast & 
\textbf{Use caching and warm-start training to speed up CI workflows} \bigstrut\\ 
\midrule
\multirow[t]{10}[8]{*}{\parbox[t]{2cm}{\textbf{Testing and Validation}}} & 
Write automated developer tests & Make your build self-testing & 
\textbf{Track and log model performance metrics on each commit} \bigstrut\\
\cmidrule{2-4}
& All tests and inspections must pass & Test in a clone of the production environment & 
\textbf{Use testing granularization and prioritization strategies (e.g., run slow tests selectively after lightweight tests pass)} \bigstrut\\
\cmidrule{2-4}

& - & - & 
\textbf{Handle non-deterministic test behavior} \bigstrut\\
\midrule
\multirow[t]{4}[4]{*}
{\parbox[t]{2cm}{\textbf{Delivery and Deployment}}} & 
- & Make it easy for anyone to get the latest executable & - \bigstrut\\
\cmidrule{2-4}

& - & Automate deployment & - \bigstrut\\
\midrule

\parbox[t][1cm][t]{2cm}{\textbf{Build Integrity and Risk Prevention}} & 
Avoid getting broken code & Everyone can see what's happening & - \bigstrut\\
\midrule

\parbox[t][0.7cm][t]{2cm}{\textbf{Data and Model Management}} & 
- & - & \parbox[t][][t]{2.8cm}{\textbf{Use model versioning and dataset versioning}} \bigstrut\\
\bottomrule
\end{tabular}
\end{table}

\subsubsection*{ML-centric CI practices definition}

\subsubsection*{\textbf{ML-specific CI practice 1:} Track and log model performance metrics on each commit (e.g., accuracy, recall, F1-score, RMSE)}

ML models evolve continuously, and even minor changes in code, data, or hyperparameters can impact model performance. To ensure that updates do not degrade model quality, it is essential to track and log key performance metrics—such as accuracy, recall, F1-score, and RMSE—at each commit. 
Maintaining detailed performance logs enables teams to detect regressions early, analyze performance trends over time, and ensure that changes align with expected quality standards.

One of the challenges identified by participants in $RQ1$ is the need for systematic model evaluation within CI pipelines. 
This aligns with the broader research challenge highlighted by Garg et al. \citep{garg2021continuous}, who emphasize the difficulty of monitoring the effectiveness of AI models over time.
As P25 explained, \textit{``there is a need to [...] track model performance metrics on each commit''}. 
Similarly, P112 explained, \textit{``Testing ML models typically involves evaluating performance metrics (e.g., accuracy, precision, recall) rather than simply checking for correct outputs''}. By integrating automated performance tracking into CI workflows, teams can establish a proactive monitoring system that ensures models meet expected quality benchmarks throughout their lifecycle.

\subsubsection*{\textbf{ML-Specific CI Practice 2:} Use testing granularization and prioritization strategies (e.g., run slow tests selectively after lightweight tests pass)}

ML projects often require large datasets, making testing resource-intensive and costly. To optimize CI  efficiency, test prioritization strategies should be implemented. For instance, tests should be classified based on their execution time and resource requirements---fast tests (e.g., unit tests) should run first, while slow tests (e.g., integration tests and model performance tracking tests) should be executed only if the lightweight tests pass.

In the results of $RQ3$, P40 suggested, \textit{``One could break the testing workflow into smaller flows [...]. Smaller unit tests are great for capturing obvious errors and they eliminate requirement to serve the whole model to capture these at some level''}.
By running complex tests only when necessary, CI workflows can minimize resource consumption while maintaining comprehensive validation of ML projects.
Such strategy not only alleviates bottlenecks in CI pipeline of ML projects, but also align with \cite{giray2021software} recommendations of scalable testing strategies tailored to the computational demands of ML workflows.

\subsubsection*{\textbf{ML-Specific CI Practice 3:} Use model versioning and dataset versioning (e.g., DVC)}

ML models rely heavily on data, and without proper dataset versioning, reproducing past results and maintaining consistency across experiments becomes challenging. Unlike traditional software projects, where code is the primary versioned artifact, ML projects require tracking datasets and trained models alongside code changes to ensure reproducibility and traceability.

By adopting dataset versioning tools such as Data Version Control (DVC), ML projects can link models to the exact datasets and code versions used during training. This improves traceability, enables experiment reproducibility, and provides rollback capabilities, allowing teams to revert to previous dataset or model versions when needed.

\subsubsection*{Examples of Dataset Versioning in CI Workflows}

To ensure efficient dataset management in CI pipelines, P112 suggested, \textit{``Implement data versioning tools like Data Version Control (DVC) to manage datasets and track changes efficiently, ensuring that only the necessary data transformations are performed during each build''}. 
This approach prevents unnecessary data reprocessing, reducing CI pipeline execution times.
Moreover, caching strategies can further enhance efficiency. P97 recommended, ``\textit{use dedicated runner machine with a proper caching and tools like DVC to minimize duplication of work between consecutive builds. Also, what could help with the latter is the CI tools like ConcourseCI or pipeline orchestrators like Dagster which have declarative pipeline execution definition based on the resource state instead of imperative plan (DVC implements similar functionality for the inner execution scope)''}.
This ensures that datasets do not need to be reloaded and reprocessed in every CI run.

\subsubsection*{Examples of Model Versioning in CI Workflows}

In addition to dataset versioning, model versioning presents another challenge that must be addressed in ML CI workflows. Large models may be infeasible to retrain fully within CI pipelines, requiring efficient model tracking and lineage management. As P57 explained,
\textit{``Our models are often too big to run end-to-end in a CI pipeline. Data and model versioning and lineage tracking are also challenging''}. 
One practical strategy, as suggested by P66, is to use a model registry:
\textit{``Try implementing model registry to maintain versions of models''}.

A model registry (e.g., \textsc{MLflow Model Registry}, \textsc{TensorFlow Model Garden}, or \textsc{SageMaker Model Registry}) provides fuctionalities such as: (i) version control for trained models, enabling rollback to previous versions if issues arise; (ii) metadata tracking, associating models with the datasets and hyperparameters used during training; and (iii) deployment automation, ensuring that only validated models are promoted to production environments. 
However, challenges persist in model versioning, as observed in pre-trained language model (PTLM) repositories on \textsc{Hugging Face}, where version identifiers often lack clear structure, and changes between versions are not well-documented~\citep{ajibode2025semverPTLM}. These inconsistencies highlight the need for standardized versioning mechanisms to enhance reproducibility and facilitate CI automation in ML projects.

By integrating dataset and model versioning into CI workflows, ML teams can achieve greater reproducibility, reduce redundant computations, and ensure that models are trained, tested, and deployed with consistent data and configurations. Future research should focus on developing standardized best practices for dataset and model versioning in CI pipelines to further improve automation and efficiency.

\subsubsection*{\textbf{ML-Specific CI Practice 4:} Handle non-deterministic test behavior}

As highlighted in the results of $RQ1$ and $RQ3$, testing ML projects presents unique challenges due to the inherent non-determinism of model behavior. ML models often produce slightly different outputs across test runs due to factors such as random weight initialization, stochastic optimization, and variations in data sampling. This randomness complicates validation, as traditional unit tests focus on code correctness, whereas ML validation depends on performance metrics like accuracy, precision, and recall. Unlike conventional software testing, where outcomes are typically binary (pass/fail), ML testing requires evaluating model performance over multiple runs, introducing additional complexity.

Recent research efforts have sought to mitigate the impact of non-determinism in ML testing. \cite{xia2023balancing} introduced FASER, a technique designed to enhance the fault-detection effectiveness of non-deterministic tests in ML projects. \textsc{FASER} systematically adjusts assertion bounds to balance the trade-off between test flakiness and fault-detection capability, thereby improving test reliability.
Additionally, \cite{rivera2021challenge} explored the sources of non-determinism in ML systems and developed \textsc{ReproduceML}, a framework aimed at promoting deterministic evaluation of ML experiments. This framework enables researchers to assess the effects of software configurations on ML training and inference, facilitating reproducibility.

To further mitigate non-determinism in ML testing, practitioners can implement several strategies, such as
fix random seeds for libraries like \textsc{NumPy}, \textsc{TensorFlow}, and \textsc{PyTorch} to ensure consistency in weight initialization and dataset shuffling, and standardizing hardware environments by running ML pipelines on identical hardware/software configurations, for instance, using containerized environments (e.g., \textsc{Docker}) and specifying exact dependency versions in \textit{requirements.txt} or \textit{conda.yml} files ensures that models are trained and evaluated in a controlled setting.
Strategies such as those could be adopted to reproducible results across different runs and environments.




\subsubsection*{\textbf{ML-Specific CI Practice 5:} Use Caching and Warm-Start Training to Speed Up CI Workflows} 

Training ML models from scratch is often infeasible in CI. Using cached computations and warm-start training (e.g., reusing previous model weights) strategies might be incoraged to reduce build times while maintaining model quality.

As P100 highlighted, \textit{``ML projects typically involve extensive data handling, including loading, cleaning, transforming, and preprocessing. These steps can be computationally intensive, especially when dealing with large datasets. This contributes to longer build durations. For example, mlpack leverages caching and parallel compilation to minimize build times. We also use automated build tools like CMake to streamline the build process''}.
In addition, P112 recommended, \textit{``Use caching mechanisms to avoid reprocessing the same datasets for every build. Store preprocessed data in a shared cache that can be reused''}. Similarly, P139 recommended, \textit{``Optimise the code across the whole ML pipeline to parallelise computations, caching when appropriate, etc''}.

Therefore, to address this challenge, CI workflows should incorporate caching mechanisms and warm-start training. By minimizing redundant computations, CI pipelines can significantly reduce build times while maintaining model performance and efficiency.

\subsection{Handling Build Duration and Test Coverage in ML Projects}
\label{sec:handling_build_duration_and_test_cov}

\subsection*{\textit{\textbf{Implication 2: Practitioners should be aware that CI pipelines for ML projects often have longer build duration expectations.}}}

Our findings from $RQ2$ indicate that practitioners' expectations regarding build durations in ML projects significantly differ from those in traditional software projects. While reducing build time is generally a key goal in CI, ML practitioners, specially from large-sized projects, recognize that longer builds are often unavoidable due to inherent complexities such as data processing, model training, and resource-intensive computations.
This aligns with our key findings in $RQ2$, where 75\% of participants expected longer builds in ML projects compared to non-ML projects, driven by the higher computational demands and dependency-heavy environments.

Furthermore, project size variability also influences acceptable build durations. Our results show that small and medium-sized ML projects tend to favor build times of 10–20 minutes, while large-sized projects are more
tolerant of longer builds, with 20.3\% of practitioners considering build times exceeding 30 minutes acceptable.
hese findings suggest that expectations of CI efficiency in ML projects should be contextualized---what is considered an excessively long build in traditional software may be seen as reasonable in ML workflows.

While it is acknowledged that CI builds in ML projects are generally longer than in non-ML projects, excessively extended build times can hinder iterative testing and experimentation, ultimately stifling the exploratory nature of ML development.
Future work should explore adaptive CI mechanisms to mitigate build times in ML workflows without compromising validation rigor. In this study, we propose and discuss testing prioritization strategies as a recommended CI practice for ML projects. However, there remains significant room for further investigation into additional strategies that can enhance CI testing efficiency in the ML domain.

\subsection*{\textit{\textbf{Implication 3: Practitioners and researchers should establish clear guidelines for managing dependencies in the CI pipeline of ML projects.}}}

Our findings for $RQ2$ highlight that dependency management is a critical factor in controlling CI build duration in ML projects. Unlike traditional software projects, ML pipelines rely on large, complex dependency trees, often requiring hardware-specific libraries (e.g., CUDA, cuDNN, TensorRT) and deep learning frameworks (e.g., TensorFlow, PyTorch). These dependencies introduce long installation times, version conflicts, and resource-intensive compilation steps, significantly increasing build duration.

Participants emphasized that ML projects tend to have extensive dependencies, many of which are hardware-specific, making installation and configuration more complex. Deep learning frameworks, widely used across various domains such as natural language processing, computer vision, and reinforcement learning, require substantial setup efforts, further extending build times. Additionally, managing GPU/CPU-specific dependencies, such as CUDA toolkit versions, presents challenges in integrating machine-specific configurations into CI pipelines, adding another layer of complexity.


Despite these challenges, current CI guidelines lack explicit recommendations for managing dependencies in ML workflows. 
Given that third-party libraries are now a fundamental component of software integration, CI practices should offer clearer guidance on dependency management, particularly for ML projects, which require a significantly larger number of dependencies ($RQ2$).

Based on the responses of the survey respondents in \textsc{Question \#3.3}, we define an initial set of strategies that should be considered when managing dependence in ML projects. These strategies are explained in the following.

\subsubsection*{Practitioners' Strategies to Optimize Dependency Management in ML CI Pipelines}

\textit{\textbf{Caching Dependencies.}}
Participants widely recommended caching dependencies to reduce redundant installations and improve CI efficiency. As P26 advised,  \textit{``Cache as many build dependencies as possible, reduce unnecessary dependencies, employ better software engineering practises''}. 
Similarly, P150 emphasized the importance of use pre-built dependencies, \textit{``Improve incremental builds use better build tools e.g. containers, nix, etc to track dependencies better. Focuse on build-once and use-many places philisophy for its dependencies''}. 

\textit{\textbf{Minimizing Unnecessary Dependencies.}}
Participants also stressed the importance of efficient dependency management, ensuring that only necessary libraries are included in each build.
As P118 recommended, \textit{``Efficiently manage dependencies and ensure that only necessary dependencies are included in each build, reducing unnecessary computations''}. 
P101 added, \textit{``As far as possible, they [ML projects] should use customized functionalities and reduce dependencies. While it's true that dependencies make coding easier, it's impossible to tell how they're affecting the build time''}. P100 further illustrated this by explaining how mlpack removed its dependency on the Boost library, significantly improving build times: \textit{``mlpack removed its dependency on the boost library relatively recently (which required refactoring most of the Neural Network codebase), and this improved our build times substantially''}.

\textit{\textbf{Containerization for Dependency Management.}}
Containerization was another frequently mentioned strategy for handling dependencies effectively. As P70 advised, \textit{``be smart about managing docker dependency changes''}. In addition, P58 suggested using lightweight containers, stating, \textit{``Simple solutions are to use tiny containers. Have something that filters out the requirements as many ML projects include excess dependencies that are only used rarely if at all''}.
In addition, P62 suggested the usage of \textit{``Small models, pre-created images with dependencies, etc''}.

To address these dependency management challenges,
future research should focus on defining standardized guidelines for dependency management in ML CI pipelines. For example, adopting containerized environments (e.g., \textsc{Docker}) with pre-installed dependencies could eliminate setup overhead. Instead of reinstalling dependencies from scratch in every run, CI workflows could pull optimized base images with the necessary libraries, potentially reducing build times and improving efficiency.

\subsection*{\textbf{\textit{Implication 4: ML teams should foster interdisciplinary collaboration and invest in testing education for ML practitioners.}}}

ML testing requires expertise that spans software engineering, data science, and domain knowledge. However, many ML developers often lack background in software testing, while software engineers may not fully understand ML-specific challenges. 
Our findings from $RQ3$ highlight that \textit{multidisciplinary and organizational} factors significantly impact the testing process of ML projects.

Organizational culture plays a crucial role in shaping test coverage expectations.
Many ML teams prioritize rapid iteration and innovation over rigorous testing, often perceiving testing as a barrier rather than an essential practice. 
As P139 noted, \textit{``ML is a fast-moving field compared to other areas. The time pressure to build and deploy ML models is high, and test coverage is simply not a priority''}.
This mindset contributes to low test coverage and a lack of structured testing strategies. To improve testing in ML projects, P135 emphasized the importance of fostering a testing culture, stating: \textit{``Start measuring it. It’s often not part of data science team’s culture to even write tests''}. Similarly, P17 highlighted the need to consider testing in project planning: \textit{``Take testing into account when providing a deadline''}.

To address these challenges, organizations should embed testing as a core component of ML workflows rather than treating it as a secondary concern. 
This involves systematically tracking test coverage, training ML developers in software testing practices, and promoting interdisciplinary collaboration between AI researchers, software engineers, and domain experts. 
For instance, when asked about strategies to enhance the test coverage of ML projects, 
P147 suggested, ML teams should \textit{``measure the test coverage and work hard to improve the coverage''}. 
In addition, P34 recommended \textit{``Improving software engineering expertise of ML practitioners''}. 
As a practical approach, P34 stressed the importance of \textit{``Onboard new engineers to software development practices''}.

Future research should establish best practices for testing ML projects, providing structured guidelines to balance the need for innovation with robust software quality assurance. 

\subsection*{\textbf{\textit{Implication 5: CI pipelines should integrate structured test coverage monitoring and enforcement mechanisms.}}}

In addition to fostering a stronger testing culture, some ML practitioners emphasized the importance of \textit{automated test coverage enforcement within CI pipelines}. As P34 suggested, CI workflows should \textit{``make code coverage a required check''}, while P56 recommended \textit{``Add a CI step that fails if test coverage drops below a certain treshold''}. 

To make test coverage monitoring more actionable, some participants emphasized the need for \textit{better visualization tools}. For example, P39 recommended using platforms like \textsc{Codecov}, while P113 suggested incorporating \textit{``Web panels, Slack bots, anything that shows coverage in an easy way, so that engineers can immediately see a problem and be motivated to fix it''}. 

Building on the importance of real-time coverage monitoring, P141 noted that CI tools should notify teams about test coverage levels, while also allowing engineers to configure acceptable coverage thresholds: \textit{``CI tools can help to notify about test coverage, but still someone needs to correct set up the values for acceptable rate''}. 

However, our results for $RQ3$ indicate that there is no consensus regarding an acceptable test coverage rate for ML projects. While most practitioners (63\%) favored coverage rates of 70–100\%, a significant number (20\%) accepted moderate coverage (50–70\%), citing ML-specific constraints such as resource demands and the stochastic nature of ML algorithms. 
These findings suggest that CI pipelines should adopt flexible, context-aware test coverage policies that balance software quality with ML-specific constraints. Additionally, the test coverage thresholds identified in $RQ3$ can serve as an initial benchmark for defining CI workflow checks in ML projects, helping teams establish realistic and effective coverage goals.

\section{Limitations}
\label{sec:limitations}


We are aware that the self-selection bias -- 
 where individuals choose whether to participate in a study -- may have influenced our results.
Out of the 5,007 practitioners we contacted via email, we received 155 responses, yielding a 3.1\% response rate. As a result, practitioners who did not respond may hold different perspectives on the presented questions, potentially leading to different findings.

However, the respondents likely represent a subset of practitioners who are more experienced or particularly interested in CI practices for ML projects. For instance, one contributor from the \textit{dmlc/tvm} project replied to the survey invitation stating, \textit{``I do not have informed opinions on CI/ML practices''}, while another contributor from the \textit{huggingface/pytorch-pretrained-BERT} project responded, \textit{``Thank you for your interest; however, I don't have any valuable input. I have no experience with ML and \textsc{GitHub}'s CI and don't want to skew your data''}. 
While this self-selection may have influenced the composition of our sample, it aligns with our study’s goal of capturing insights from practitioners who have experience with CI in ML projects. Their perspectives provide valuable feedback on the key differences, challenges, and strategies for adopting CI in this context.

Furthermore, participants who did respond may have been affected by social desirability bias -- the tendency to answer questions in a way they believe will be viewed favorably by others. This could include providing responses perceived as ``correct'' or portraying their development practices more positively than they actually are. To mitigate these limitations and enhance the reliability of the findings, the study included a diverse set of ML projects across different build durations, ensuring representation from both high-performing and low-performing projects. Additionally, the qualitative analysis was conducted meticulously to identify recurring themes, reducing the impact of potential biases and strengthening the robustness of the results.

The study focuses on 47 ML projects, targeting those in the top and bottom quartiles for build durations. While this approach captures contrasting experiences and highlights key challenges, it may exclude insights from projects with average build durations, potentially narrowing the scope of findings. However, by emphasizing the extremes, the study provides a richer understanding of the factors influencing build durations. Future research could include projects with average durations to offer a more comprehensive perspective.

Additionally, the analysis of small ML projects is based on only two responses, which significantly limits the generalizability of conclusions for this group. Findings related to small projects are explicitly discussed with caution, acknowledging their limited reliability. These insights are framed as exploratory and are intended to provide a foundation for future investigations rather than definitive conclusions.

The study also focuses exclusively on ML projects using \textsc{GitHub Actions} as their CI platform. The challenges and insights identified may differ for projects using alternative CI tools or operating in varied organizational or technological contexts. For instance, some findings, such as resource and infrastructure challenges, may reflect specific limitations of \textsc{GitHub Actions} (e.g., GPU availability). A broader analysis that includes multiple CI platforms could uncover additional challenges and strategies to address them.

Finally, the thematic analysis employed in the study relies on the manual coding of open-ended survey responses, which is an inherently subjective process, introducing the potential for varying interpretations among authors. To mitigate this threat, two authors coded each open-ended survey question independently, and disagreements were resolved by a third author. Although we recognize that alternative interpretations of the data could exist, we believe this triangulation approach reduces individual biases and ensures consistency and reliability in the coding process.

\section{Conclusion}
\label{sec:conclusion}

Our study qualitatively investigated practitioners' perspectives on the adoption of CI practices in ML projects, drawing insights from a survey of 155 participants. 
By exploring key challenges, such as build duration management, test coverage expectations, and reasons behind the differences between ML and non-ML when adopting CI practices, the study provides a comprehensive understanding of the unique demands of ML projects in the context of CI.

The key takeaway from this study is that while foundational CI principles remain valuable, ML projects require tailored approaches to address their unique challenges, including longer build durations, lower test coverage, complex dependency management, and the non-deterministic nature of ML workflows.
To bridge this gap, we propose a set of ML-specific CI practices, such as tracking model performance metrics on each commit, prioritizing test execution, and versioning datasets and models. Additionally, our study underscores the importance of fostering interdisciplinary collaboration to strengthen the testing culture in ML projects. Finally, our discussion highlights the need for standardized CI guidelines, particularly in dependency management, to improve CI effectiveness in ML workflows.
This study identifies critical gaps between current CI practices and practitioner expectations and provides a roadmap for advancing CI in ML projects through targeted research and practical interventions.

\section*{Data Availability Statements}

The replication package for this study, including anonymized datasets, survey responses, and analysis scripts, is available at \url{https://zenodo.org/records/14902811}.
Due to privacy concerns, raw survey responses containing non-anonymized text are not publicly available.

\section*{Funding}

This work is partially supported by \textsc{INES} (National Institute of Software Engineering), \textsc{CNPq} grant 465614/2014-0, \textsc{CAPES} grant 88887.136410/2017-00, \textsc{FACEPE} grants APQ-0399-1.03/17, \textsc{PRONEX} APQ/0388-1.03/14, and \textsc{CNPq} grant 425211/2018-5.

\section*{Author Contributions}

\begin{itemize}
    \item João Helis Bernardo: Conceptualization, Methodology, Data Collection, Thematic Analysis, Writing—Original Draft.
    \item Daniel A. da Costa: Methodology, Writing—Review \& Editing, Thematic Analysis, Supervision, Research Direction.
    \item Filipe R. Cogo: Methodology, Writing—Review \& Editing, Thematic Analysis.
    \item Sérgio Q. de Medeiros: Methodology, Writing—Review \& Editing, Thematic Analysis.
    \item Uirá Kulesza: Methodology, Writing—Review \& Editing, Thematic Analysis, Supervision, Research Direction.
\end{itemize}

\section*{Ethical Approval}

The study involved a survey with software practitioners regarding their experiences with continuous integration in machine learning projects. The research did not involve personal or sensitive data and adhered to ethical guidelines for research with human participants. 
The study did not collect personally identifiable information beyond publicly available \textsc{GitHub} information. Furthermore, participants were informed about the research purpose, and their responses were anonymized.

\section*{Informed Consent}

The invitation email included a description of the study, its purpose, and the voluntary nature of participation. No personally identifiable information was collected beyond publicly available \textsc{GitHub} information.

\section*{Conflict of interest}

The authors declare that they have no conflicts of interest related to this work.

\bibliographystyle{spbasic}      
	
\bibliography{main}

\appendix
	
\appendixtitleon
\appendixtitletocon
\begin{appendices}

\section{Invitation Email Example}
\label{sec:appendix_invitation_email_example}

\begin{tcolorbox}[colframe=black!75, colback=white!95, title=SUBJECT - Help us understand CI in ML: Your Experience is Essential,
fonttitle=\normalsize]
\normalsize

Dear \textbf{participant\_name}, \\

I hope you're well. \\

We are a group of researchers from universities based in Brazil, New Zealand, Canada, and France on a study about Continuous Integration (CI) practices in Software Projects using Machine Learning (ML projects). \\

We'd love your input as you have successfully submitted \textbf{{ pull\_contributions\_number }} pull request that was/were merged on the \textit{main/master} branch of \textbf{{ project\_full\_name }} project since it adopted GitHub Actions. \\

Please click the link to participate: \textbf{{ google\_form\_link}} \\

The survey has 8 open and 12 closed questions and takes \textbf{less than 10
minutes}. Your responses are confidential and will help us better understand CI practices in ML projects. \\

As a thank you, all participants who complete the survey will enter 
a draw for \textbf{ten \$50 Amazon or Steam gift cards}. \\

Thank you for your time and insights! \\

Best regards, \\

\textit{Joao Helis Bernardo} \\
\textit{Federal University of Rio Grande do Norte, Brazil}

\end{tcolorbox}

\end{appendices}

\end{document}